\documentclass[prd,nofootinbib,preprint,superscriptaddress]{revtex4}

\usepackage{amsmath, amssymb, amsthm, graphicx, epsfig, fancyhdr,epsfig, slashed}
\usepackage{tikzsymbols}
\usepackage{natbib}
\usepackage{float}

\usepackage{mathtools} 

\usepackage{tikz,xcolor,hyperref}

\definecolor{lime}{HTML}{A6CE39}
\DeclareRobustCommand{\orcidicon}{
	\begin{tikzpicture}
	\draw[lime, fill=lime] (0,0) 
	circle [radius=0.2] 
	node[white] {{\fontfamily{qag}\selectfont \tiny ID}};
	\draw[white, fill=white] (-0.0625,0.095) 
	circle [radius=0.007];
	\end{tikzpicture}
	\hspace{-2mm}
}

\foreach \x in {A, ..., Z}{\expandafter\xdef\csname orcid\x\endcsname{\noexpand\href{https://orcid.org/\csname orcidauthor\x\endcsname}
			{\noexpand\orcidicon}}
}

\newcommand{\be}{\begin{equation}}
\newcommand{\ee}{\end{equation}}
\newcommand{\bea}{\begin{eqnarray}}
\newcommand{\eea}{\end{eqnarray}}

\newcommand{\spp}[1]{\textcolor{red}{[Shila: #1]}}
\newcommand{\ag}[1]{\textcolor{blue}{[Anish: #1]}}

\usepackage{ulem,soul, xcolor, color,cleveref,subcaption, multirow,ulem,cancel, footmisc,dsserif}
\hypersetup{colorlinks,linkcolor={green!50!black},citecolor={cyan},urlcolor={red}}

\newcommand{\shilacred}{\color{red!70!green}}
\newcommand{\shila}[1]{\begin{align}#1\end{align}}
\newcommand{\bi}{\bibitem}
\newcommand{\ba}{\begin{eqnarray}}
\newcommand{\ea}{\end{eqnarray}}

\newcommand{\bc}{\begin{cases}}
\newcommand{\ec}{\end{cases}}

\newcommand{\no}{\nonumber}

\newcommand{\lt}{\left }
\newcommand{\rt}{\right }
\newcommand{\seq}{\simeq}
\newcommand{\gsim}{\gtrsim}
\newcommand{\lsim}{\lesssim}

\newcommand{\mph}{m_\Phi}
\newcommand{\mvph}{m_\varphi}

\definecolor{g}{rgb}{0.1, 0.6, 0.5}

\newcommand{\stbl}{ \ttfamily \color{blue!30!red} }

\definecolor{green(html/cssgreen)}{rgb}{0.0, 0.5, 0.0}

\newcommand{\di}{i} 
\def\vp{\varphi}

\def\p{\pi}

\def\Ph{\Phi}

\def\l{\lambda}
\def\c{\chi}
\def\r{\rho}
\def\G{\Gamma}
\def\g{\gamma}
\def\gs{g_{\star}}
\def\gss{g_{\star, s}}

\def\a{\alpha}

\def\b{\beta}

\def\G{\Gamma}

\def\p{\pi}
\def\r{\rho}
\def\z{\zeta}
\def\t{\theta}
\def\T{\Theta}

\def\vr{\varrho}

\definecolor{dukeblue}{rgb}{0.0, 0.0, 0.61}

\newcommand{\shilaD}{ \color{dukeblue}}

\def\ha#1{\href{http://arxiv.org/abs/#1}{\shilacred \ttfamily #1}}

\def\hds#1{ \href{https://doi.org/#1}}
\def\jn#1{{\shilaD #1}}

\newcommand{\Planck}{\textit{Planck}}

\newcommand{\KeckArray}{\textit{Keck Array}}
\newcommand{\BICEP}{\textsc{Bicep}}

\def\ss{\mathbb{s}}
\def\sA{\mathbb{A}}
\def\mH{m_H}
\def\mc{m_{\chi}}

\def\mp{\, M_P}
\def\lH{\l_H}

\def\yc{y_\c}
\def\Yc{Y_\c}
\def\Yco{Y_{\c,0}}
\def\Yzo{Y_{\z,0}}
\def\yz{y_\zeta}
\def\lO{\l_{12}}
\def\lT{\l_{22}}
\def\cc{\bar{\chi}\chi}
\def\ccp{\chi\chi}
\def\HH{H^\dagger H}
\def\UP{U_\Phi(\Phi)}
\def\PhC{\Phi_{\rm CMB}}

\def\PhZ{\Phi_{0}} 
\def\vpZ{\varphi_{0}}
\def\uP{U_\Phi}
\def\duP{U'_\Phi}
\def\dduP{U''_\Phi}

\def\uvp{U_\varphi}
\def\duvp{U'_\varphi}
\def\dduvp{U''_\varphi}

\def\Uc{U_\chi(\chi)}
\def\UH{U_H(H)}
\def\Trh{T_{rh}}
\def\Tmax{T_{max}}
\def\tw{\text{w}}

\def\td{\text{d}}

\def\Br{\text{Br}}

\def\tl{\theta_l}
\def\zz{\z^\dagger\z}
\def\rr{\vr\vr}

\def\in{{\rm in}}

\def\hh{h h}

\def\sm{SM}
\def\dm{DM}
\def\cdm{CDM}
\def\sm{SM}
\def\cmb{CMB}
\def\rd{RD}
\def\md{MD}
\def\sdm{SDM}
\def\bbn{BBN}

\def\mdl{\text{Model} }

\def\fmI{\qquad  \text{(for Model I)} }
\def\fmII{\qquad  \text{(for Model II)} }
\def\fmIt{\text{(for Model I)} }
\def\fmIIt{\text{(for Model II)} }

\hypersetup{colorlinks,linkcolor={green!70!blue},citecolor={magenta},urlcolor={magenta}}
\definecolor{DGreen}{rgb}{0.68, 1.00, 0.18}

\definecolor{dp}{rgb}{1.0, 0.8, 0.64}

\begin{document}


\title{\textcolor{black}{\bf Inflection-point Inflation and Dark Matter Redux}}

\author{Anish Ghoshal\orcidA{}}
\email{anish.ghoshal@fuw.edu.pl}
\affiliation{Institute of Theoretical Physics, Faculty of Physics, University of Warsaw, ul. Pasteura 5, 02-093 Warsaw, Poland}

\author{Gaetano Lambiase\orcidG{}}
\email{lambiase@sa.infn.it}
\affiliation{Dipartimento di Fisica ”E.R. Caianiello”, Universita’ di Salerno, I-84084 Fisciano (Sa), Italy}
\affiliation{INFN - Gruppo Collegato di Salerno, I-84084 Fisciano (Sa), Italy}


\author{\\ Supratik Pal\orcidP{}}
\email{supratik@isical.ac.in}
\affiliation{Physics and Applied Mathematics Unit, Indian Statistical Institute, Kolkata-700108, India}
\affiliation{Technology Innovation Hub on Data Science, Big Data Analytics and Data Curation,
Indian Statistical Institute, Kolkata-700108, India}

\author{Arnab Paul\orcidC{}}
\email{arnabpaul9292@gmail.com}
\affiliation{Physics and Applied Mathematics Unit, Indian Statistical Institute, Kolkata-700108, India}
\affiliation{School of Physical Sciences, Indian Association for the Cultivation of Science, Kolkata-700032, India}

\author{Shiladitya Porey\orcidS{}}
\email{shiladityamailbox@gmail.com}
\affiliation{
Department of Physics, Novosibirsk State University, Pirogova 2, 630090 Novosibirsk, Russia}

\begin{abstract}

\textit{We investigate for viable models of inflation that can successfully produce dark matter (DM) from inflaton decay process, satisfying all the constraints from Cosmic Microwave Background (CMB) and from some other observations. 
In particular, we analyze 
near-inflection-point small field inflationary scenario with non-thermal production of fermionic DM 
from the decaying inflaton field during the reheating era. To this end, we propose two different models of inflation with polynomial potential.
The potential of \mdl I contains terms proportional to linear, quadratic, and quartic in inflaton; whereas in \mdl II, the potential contains only even power of inflaton and the highest term is sextic in inflaton. For both the models, we find out possible constraints on the model parameters 
which lead to proper inflationary parameters from CMB data
with a very small tensor-to-scalar ratio, 
as expected from a small-field model. 
With the allowed parameter space from CMB, we then search for
satisfactory relic abundance for DM, that can be produced from inflaton via reheating,
 to match with the present-day cold dark matter (CDM) relic density for the parameter spaces of the DM $\chi$ mass and Yukawa couplings in the range $10^{-9} \gsim \yc \gsim 10^{-15}$ and $10^3 \text{GeV} \lsim \mc \lsim 10^9 \text{GeV}$. The DM relic is associated with the inflection-points in each model via maximum temperature reached in the early universe during its production.
 Finally, we find out allowed parameter space coming out of combined constraints from stability analysis for  both SM Higgs and DM decays from inflaton as well as from BBN and Lyman-$\alpha$ bounds.} 

\end{abstract}

\maketitle

\section{Introduction}
Cosmic inflation, as originally proposed in Refs.~\cite{Starobinsky:1980te, Guth:1980zm, Linde:1981mu, Albrecht:1982wi}, postulated a brief period of non-adiabatic exponential expansion of space-time leading to washing out of all imprinted primordial information, including preexisting inhomogeneities in energy-momentum tensor and spacetime metric. This explains the high degree of spatial flatness, homogeneity, and
isotropic nature of the universe, and uniform (2.725 K) black body radiation as seen in the temperature fluctuations of Cosmic Microwave Background Radiation (CMBR) measurements. During inflation, the Hubble parameter is nearly constant indicating quasi de-sitter universe, that generically produces scalar and tensor quantum fluctuations characterized by nearly scale invariant power spectrum on super-horizon scales, which upon re-entering at a much later stage of evolution of the universe, produces large scale structures (LSS), primordial gravitational waves (PGW) and may also source primordial black holes (PBHs). Properties of these primordial scalar and tensor perturbations, amplitude of scalar power spectrum $A_s$, scalar spectral index $n_s$, and tensor-to-scalar ratio $r$ from CMB measurements are obtained from \Planck~and \BICEP~observations~\cite{Planck:2018jri, BICEP2:2018kqh}. From the measured values of $n_s$ and $r$, $V(\phi) \propto \phi^p$  with $p \geq 1$ inflationary potentials have been disfavored (see e.g., Ref.~\cite{Martin:2013tda} for a comprehensive review). However, plateau-type potentials are compatible with current CMB predictions.

Besides addressing inflation, the $6$-parameter ($\Lambda$CDM) fit of the \cmb~data manifests that $\sim 26.5\%$ of the total mass-energy density of the universe is in the form of Cold Dark Matter (\cdm) while the visible matter is only $\sim 5\%$. However,  even though we have several gravitational signatures of DM on top of \cmb~data, there is no direct non-gravitational signature of DM so far. So, the exact nature of DM as well as its origin in the universe still eludes us. 

There are numerous proposed models for DM particles originating at different epochs and via various production mechanisms~\cite{Bernal:2018hjm} in the history of the universe%
. Their cross-section with other particles, and production channels control the genesis of those particles and time of decoupling from the plasma of the universe. In this work, we are considering one of the earliest possible epochs of production of \dm~i.e. during reheating.~%
 The most popular model of non-thermal DM is known as the Weakly Interacting Massive Particles (WIMP) scenario. However, lack of direct evidence of the existence of WIMP particles in the colliders, make this scenario questionable. If the DM particles are so feebly interacting with the visible sector particles that they never reach thermal equilibrium with the relativistic SM plasma, then the so-called feebly interacting massive particle (FIMP)~\cite{McDonald:2001vt, Choi:2005vq, Kusenko:2006rh, Petraki:2007gq, Hall:2009bx, Bernal:2017kxu, Hall:2009bx} are produced from interaction with the thermal bath. To say more specifically, when the temperature of the thermal bath drops below the mass of  DM particles, the abundance of DM is  frozen-in. A notable difference between FIMP and WIMP is that DM abundance via the former mechanism depends on the initial conditions, namely, the initial density (typically assumed to be small) of DM particles, which can be produced during the cosmic re-heating stages. 
Besides, there remains another possibility for the DM genesis which involves the direct decay of a heavy particle, for example, a moduli field, the curvaton (see, e.g. Ref.~\cite{Baer:2014eja}), or the inflaton which is contemplated in this work. Additionally, DM particles could have also been produced via scattering of SM particles or inflatons mediated by the irreducible gravitational interaction~\cite{Garny:2015sjg, Tang:2016vch, Tang:2017hvq, Garny:2017kha, Bernal:2018qlk}.

	In this paper, we look to have a single unified model of inflation and the production of dark matter. Inflaton as the SM gauge singlet are studied in several works ~\cite{Lerner:2009xg, Kahlhoefer:2015jma}), and particularly with non-minimal coupling between inflaton and the Ricci scalar Refs.~\cite{Clark:2009dc, Khoze:2013uia, Almeida:2018oid, Bernal:2018hjm, Aravind:2015xst, Ballesteros:2016xej, Borah:2018rca,  Hamada:2014xka, Choubey:2017hsq, Cline:2020mdt, Tenkanen:2016twd, Abe:2020ldj}
	flattens the potential in the Einstein frame. Having flat potentials without such non-minimal coupling to gravity have explored in BSM model scenarios like the $\nu$MSM~\cite{Shaposhnikov:2006xi}, the NMSM~\cite{Davoudiasl:2004be}, SMART $U(1)_X$ \cite{Okada:2020cvq}, the  WIMPflation~\cite{Hooper:2018buz}, model with a single axion-like particle~\cite{Daido:2017wwb}, and extension with a complex flavon  field \cite{Ema:2016ops}\footnote{ For supersymmetric models, see, e.g., Ref.~\cite{Allahverdi:2007wt}, which is inspired by the gauge-invariant MSSM inflation~\cite{Allahverdi:2006iq}.}. Recently Ref.~\cite{Ghoshal:2022ruy} looked at inflaton decay to dark matter with complementary DM versus gravitational searches.
	
	Particularly inflection-point inflation and creation of inflection point in various particle physics models have been of great interest \cite{Ghoshal:2022zwu,Ghoshal:2022hyc,Ballesteros:2015noa,Senoguz:2008nok,Enqvist:2013eua,Okada:2017cvy,Choi:2016eif,Allahverdi:2006iq,BuenoSanchez:2006rze,Baumann:2007np,Baumann:2007ah,Badziak:2008gv,Enqvist:2010vd,Cerezo:2012ub,Choudhury:2013jya,Choudhury:2014kma,Okada:2016ssd,Stewart:1996ey,Stewart:1997wg,Drees:2021wgd} since it produces flat potential in a very simple framework and can be embedded quite easily in BSM models. It may avoid trans-Planckian issues~\cite{Bedroya:2019tba} and is consistent with the swampland distance conjecture~\cite{Agrawal:2018own}, due to its small field inflation scenario. Usually the running of the spectral index, $\alpha_s \sim \mathcal{O}(10^{-3})$, also falls within the testable regions in the foreseeable future~\cite{Munoz:2016owz}. Inflection-point inflation can lead to the scale of inflation $H_I$ to be very low, and therefore no cosmological moduli problem in the universe exists in this set-up~\cite{Coughlan:1983ci}.
	
	
With such motivations for creating inflection-point to drive inflation in the early universe, we consider an additional gauge singlet fermion field $\chi$ as the DM candidate ~\cite{Kim:2006af,Kim:2008pp, Baek:2011aa, Lopez-Honorez:2012tov, Kim:2014kok, Choi:2020ara}~
together with the scalar singlet as the inflaton. Along with that, we assume no or very tiny mixing between the infaton and the SM Higgs. 

	Due to its direct coupling with the DM, the inflaton either decays to DM or may undergo scattering with the dark sector to produce the observed relic. As we will see, additional irreducible gravitational interaction may also mediate the DM production, either by 2-to-2 annihilation of the SM Higgs bosons, or of the inflatons during the reheating era.
	
	The paper is organized as follows: In Section~\ref{Sec:Inflection-point Inflation Models}, we discuss the definition of inflection point for a single field potential. In Subsection~\ref{sec:Inflection-point achieved with Linear term}, we explore the slow roll inflationary scenario of \mdl I inflation and also, fix the coefficients of the inflaton potential from the \cmb~data. The inflationary scenario, together with the estimation of the coefficient of the inflaton potential of \mdl II, are discussed in Subsection~\ref{sec:Inflection-point achieved with Sextic Term}. In the succeeding subsection (Subsection~\ref{sec:Stability analysis for linear term inflation}), we analyzed the stability of the inflaton-potential of both models with respect to radiative correction. In Section~\ref{Sec:Reheating}, we comparatively discuss the reheating scenario for both models. We also discuss whether the non-thermal vector-like fermionic \dm~produced via the decay of inflaton or from the scattering of inflaton/\sm~particles can explain the total \cdm~density of the present universe. We finally conclude in Section~\ref{Sec:Conclusions and Discussion}.

\medskip

\section{Inflection-point Inflation Models}
\label{Sec:Inflection-point Inflation Models}

The presence of an inflection point in the inflationary potential is significant for the slow roll inflation, as it provides  a plateau-like region in the potential. This plateau-like region helps to 
increase the number of e-foldings (defined in Section~\ref{sec: slow roll parameters}) without  noticeable change in the inflaton-value, while inflaton is in the neighborhood of inflection point~\cite{Okada:2016ssd, Garcia-Bellido:2017mdw}.%
\footnote{
The presence of an inflection point in the potential
can also cause a rise of the scalar power spectrum if inflaton starts the journey significantly before inflection point.
}

To find out the stationary inflection point of an inflationary potential ${\cal V}(\theta)$ of a single scalar field $\theta$, we need the solution of 
\ba \label{eq:condition_for_inflection_point}
\frac{\td {\cal V}}{\td \theta} = \frac{\td^2 {\cal V}}{\td \theta^2}=0   \,.
\ea 
Also, it is necessary that $\td^2{\cal V}/ \td \theta^2$ should change sign at the inflection point.%
\footnote{
{\it For further details, see Section3.6.1.3-1 
of Bronshtein, Ilía Nikolaevich, et al. Handbook of mathematics. Springer, 5th ed., 2007, doi:10.1007/978-3-540-72122-2.}
}
In the following two subsections (Section~\ref{sec:Inflection-point achieved with Linear term} and Section~\ref{sec:Inflection-point achieved with Sextic Term}) we discuss two different slow roll small-field inflationary scenarios, where each of the inflationary potentials possesses a near inflection point.

\subsection{Inflection-point achieved with Linear term}
\label{sec:Inflection-point achieved with Linear term}
The Lagrangian density of Model I inflation 
where the inflaton $\Ph$ is 
minimally coupled to gravity
(with $\hbar=c=k_{B}=1$) is: 
\shila{
\label{Eq:Lagrangian density-ModI}
\mathcal{L}_{I} &= \frac{\mp^2}{2} \mathcal{R}
+\mathcal{L}_{INF} (\Ph)
+ {\cal L}_{KE, \chi} 
  - U_\chi(\chi) 
  + {\cal L}_{KE, H} 
  - U_H(H)
 + \mathcal{L}_{reh, I} 
 + \mathcal{L}_{\sm}  \,,
}
where ${\cal R}$
is the Ricci scalar with metric-signature $(+,-,-,-)$ and $M_P\simeq 2.4 \times 10^{18}\text{GeV}$ is the reduced Planck mass. 
The second term on the right-hand side of Eq.~\eqref{Eq:Lagrangian density-ModI}, $\mathcal{L}_{INF} (\Ph)$, is the Lagrangian density of $\Phi$ (a real scalar field) -
\ba
&\mathcal{L}_{INF} (\Ph) = \frac{1}{2}\partial_\mu \Phi\partial^\mu\Phi - U_\Phi(\Phi) \,,\\
&\UP = V_0 + a \, \Ph - b \, \Ph^2 + d \, \Ph^4  \label{eq:inflation potential of model I} \,,
\ea
where $V_0$, $a$, $b$, and $d$ are all assumed to be positive, real, and with mass dimensions $\mp^{4}, \mp^3,\mp^2$, and $\mp^0$, respectively. The minus sign in front of $b$ helps to get the inflection point. 
To make the potential bounded from below, we choose $d > 0$. 
Such linear terms in the potential can be motivated by many instances, e.g., from non-perturbative generation of condensates under dark interaction \cite{Barenboim:2008ds,Bardeen:1989ds,Bhatt:2008hr,Bhatt:2009wb,Barenboim:2010nm}, or from gravitational interactions~\cite{Barenboim:2010db,Dvali:2016uhn}, or from SUSY motivations~\cite{Rehman:2009nq,Dvali:1994ms,Copeland:1994vg,Lazarides:2001zd,Lyth:1998xn,Takahashi:2010ky}~and several others \cite{Davoudiasl:2020opf,Iso:2014gka,Kaneta:2017lnj}. Here we do not go into such detailed models but instead, use it as a phenomenological model where the analysis remains applicable to all such scenarios. We will fix these coefficients from the constraints on the parameters of inflationary scenario, obtained from \cmb~data. In Eq.~\eqref{Eq:Lagrangian density-ModI},
${\cal L}_{KE, \chi}$, and ${\cal L}_{KE, H}$ symbolizes kinetic energy of the vector-like fermionic \dm, $\chi$, and \sm~Higgs field, $H$, respectively. The potential term for $\c$ and $H$ are given by respectively - 
\ba
&&\Uc =  \mc \cc  \, ,\\
&&\UH = -\mH^2 \HH  + \lH \lt( \HH \rt)^2 \,.
\ea
$\mathcal{L}_{\sm}$ is the Lagrangian density of the \sm~members and $\mathcal{L}_{reh, I} $ which plays a crucial role during reheating, takes care of the interactions of $\c$ and $H$ with $\Ph$~\cite{Drees:2021wgd}: 
\ba\label{eq:reheating lagrangian for modelI}
\mathcal{L}_{reh,I} = - \yc \Ph \cc - \lO \Ph \HH - \lT \Ph^2 \HH \,.
\ea 

Here $\lO$, $\lT$, and Yukawa-like $y_\c$ are the couplings of \sm~Higgs and fermionic \dm~with inflaton.
$\yc$ and $
\lT$ are dimensionless, while $\lO$ has the dimension of $\mp$.

During inflation, contribution from the other terms, except the first two terms on the right side of Eq.~\eqref{Eq:Lagrangian density-ModI}, is negligible. 
Then, the equation of motion of the spatially homogeneous and isotropic $\Ph$ is
\shila{\label{eq:Klein–Gordon equation}
\ddot{\Ph} + 3 \, {\cal H} \, \dot{\Ph} + \duP =0\,,
}
where $\mathcal{H}$ is the Hubble parameter. In Eq.~\eqref{eq:Klein–Gordon equation}, over-dot indicates derivative with respect to time $t$, and prime denotes derivative with respect to inflaton. For slow roll inflation\footnote{
In Ref.~\cite{Germani:2017bcs}, it has been shown that near the inflection point of the potential, an ultra slow-roll phase exists, and the slow roll conditions are not applicable there. However, in this work, $\PhC<\PhZ$ and inflaton does not cross the inflection point during the course of its evolution when the perturbations corresponding to the scales of interest leave the Hubble horizon. Although a choice of $\dot{\Phi}_{(\text{at } \Phi=\PhC)}=0$ as initial condition can violate the geometric slow-roll condition, choosing a small initial inflaton velocity at $\Phi=\PhC$ or starting the evolution at $\Phi>\PhC$ can resolve this issue. So, using the slow-roll approximation in this work is reasonable.
} by $\Ph$, $\ddot{\Ph}$ must be negligible so that%
\shila{\label{eq:equation of slow roll inflation}
3\,  {\cal H} \, \dot{\Ph} =- \duP \,. 
}
In addition to that, we need $\duP \ll \uP$~\cite{Lyth:1993eu, Rydan} which again implies that 
the kinetic term of the inflaton $\dot{\Phi}^2 \ll \UP$. As a result of that, the first Friedman equation leads to
\be\label{eq:Hslowroll}
\mathcal{H}^2= \frac{1}{3\mp^2 } \UP \,.
\ee
Therefore, ${\cal H}$ is nearly constant during the slow roll phase of inflation. 
 Again, Eq.~\eqref{eq:equation of slow roll inflation}, together with $\dot{\Phi}^2 \ll \UP$, suggests~\cite{Riotto:2002yw}
\begin{eqnarray}\label{eq:slow roll Hubble}
\mathcal{H}^2 \gg U_\Phi^{''}\,.
\end{eqnarray}
These conditions are generally demonstrated in terms of slow roll parameters, mentioned in the succeeding part of this text.

\subsubsection{Slow roll parameters and number of e-foldings}
\label{sec: slow roll parameters}
The slow-roll condition during inflation is measured in terms of slow-roll parameters ($\epsilon_V,\eta_V, \xi_V, \sigma_V$) whose absolute values must be $\ll 1$ throughout the inflation. 
The four slow roll parameters are respectively~\cite{Lyth:2009zz, poly,Ghoshal:2022hyc} (we have used the symbols from~\cite{Lyth:2009zz})
\bea
\epsilon_V &&\approx \frac{\mp^2}{2}\left( \frac{U_\Phi^\prime}{U_\Phi}\right)^2
=\mp^2\frac{\left(a-2 b \, \Phi +4 d \, \Phi ^3\right)^2}{2 \left(\Phi  \left(a-b \, \Phi +d \,  \Phi^3\right)+V_0\right){}^2}   \,,
\label{eq:epsilonV}
\\
\eta_V &&\approx \mp^2\frac{U_\Phi^{\prime \prime}}{U_\Phi} 
= -\mp^2\frac{2 \left(b-6 d \, \Phi ^2\right)}{\Phi  \left(a-b \, \Phi +d \, \Phi ^3\right)+V_0}   \,,
\label{eq:etaV}\\
%
%
\xi_V&&
\approx \mp^4\frac{U_\Phi^\prime U_\Phi^{\prime \prime \prime}}{U_\Phi^2} 
= \mp^4\frac{24 d \, \Phi  \left(a-2 b \, \Phi +4 d \, \Phi ^3\right)}{\left(\Phi  \left(a-b \, \Phi +d \, \Phi^3\right)+V_0\right){}^2}  \,,
\label{eq:xiV}
\\
\sigma_V &&\approx \mp^6\frac{{U_\Phi^\prime}^2 U_\Phi^{\prime \prime \prime \prime}}{U_\Phi^3} = \mp^6\frac{24 d \, \left(a-2 b \, \Phi +4 d \, \Phi ^3\right)^2}{\left(\Phi  \left(a-b \, \Phi +d \, \Phi^3\right)+V_0\right){}^3}  \,.
 \eea

As soon as the slow roll conditions are violated at $\Phi \sim \Ph_{\rm end}$, i.e. any one of the slow-roll parameters $\left|\epsilon_V\right|,\left|\eta_V\right|,\left|\xi_V\right|,\left|\sigma_V\right| \sim 1$, inflation ends. The duration of inflation is measured in terms of the total number of e-foldings, $\mathcal{N}_{\rm CMB, \, tot}$:~\cite{Lyth:2009zz,poly, Liddle:2000cg,Ghoshal:2022hyc})
\be\label{eq:def-e-fold}
\mathcal{N}_{\rm CMB, \, tot}=\mp^{-2}\int_{\Ph_{\rm end}}^{\Ph_{\rm CMB}} \frac{\uP}{\duP} \, \td \Ph =\int_{\Ph_{\rm end}}^{\Ph_{\rm CMB}} \frac{1}{\sqrt{2 \epsilon_V}} \, \td \Ph  \,, 
\ee
where $\PhC$ is the value of the inflaton at which the length scale that left the causal horizon during inflation, re-enters during the era of recombination, i.e. during the generation of \cmb.

\subsubsection{{Scalar and} Comoving curvature power spectrum and scalar spectral index}
\label{sec: comving curvature power spectrum}
From the discussion of Section~\ref{sec:Inflection-point achieved with Linear term}, we can claim that the size of the causal horizon (which is $\sim {\cal H}^{-1}$) remains nearly constant during slow roll inflation. Howbeit, any physical length scale in our expanding universe always continues to increase as $\propto \ss$, where $\ss(t)$ is the cosmological scale factor. Thus, '$k$'-th Fourier modes of quantum fluctuations, generated during inflation, may leave the causal region before the ending of inflation whenever $k<{\cal H}$. Then, on super-horizon scales, the spectrum of these Fourier modes become scale-invariant and frozen with the amplitude of the fluctuations almost independent of the time of horizon crossing. 
After the completion of inflation, ${\cal H}(t)$ changes as ${\cal H}\sim \ss^{n}$ ($n=-2$ for radiation domination and $-3/2$ for matter domination stage). Then, that '$k$'-th Fourier modes of the fluctuations may reenter the causal region. The statistical nature of these fluctuations is expressed in terms of power spectrum, which can be calculated from two point correlation function of the fluctuations. Here, it is assumed that fluctuations are Gaussian in nature~\cite{Riotto:2002yw, Baumann:2009ds}. 
The scalar power spectrum is then defined as
~\cite{Hotchkiss:2011gz, Chatterjee:2014hna, Chatterjee:2017hru}:
\be
\mathcal{P}_s \left( k \right) = A_s \left(  \frac{k}{k_*} \right)^{n_s -1 + (1/2) \alpha_s \ln(k/k_*) + (1/6)\beta_s (\ln(k/k_*))^2 }  \label{eq:define scalar power spectrum}\,,
\ee
where $A_s$ is the normalization
\bea\label{eq:As}
A_s \approx \frac{U_\Ph}{24 \pi^2  \mp^4\, \epsilon_V} \approx \frac{2 U_\Ph}{3 \pi^2 \mp^4 \, r} \,,
\eea 

and $k_*$ is the pivot scale, which is $0.05 \text{Mpc}^{-1}$ for \Planck~
observation (see Section~\ref{sec:Planck Data}). Similarly, the power spectrum for tensor perturbation is $\mathcal{P}_h \left( k \right)$ (with $A_t$ as the normalization and $n_t$ as the tensor spectral index)~\cite{Chatterjee:2014hna} 
\ba 
\mathcal{P}_h \left( k \right) = A_t \left(  \frac{k}{k_*} \right)^{n_t + (1/2) d n_t/d \ln k \ln(k/k_*) + \cdots } \,,
\ea 
and then the ratio of the amplitudes of the tensor to scalar power spectrum is~\cite{Chatterjee:2014hna, Linde:2007fr}
\be
r = \frac{A_t}{A_s}\approx 16 \epsilon_V  \,.\label{eq:r aprrox EtaV}
\ee
For adiabatic density perturbation, the comoving curvature perturbation in super-horizon limit is
\be
\Delta_\mathcal{R}^2= \left(\frac{5+ 3 w}{3\left( 1+ w\right)} \right)^2 \mathcal{P}_s  \,.
\ee
For the radiation (\rd) and matter domination (\md) stage, equation of state parameter $w =1/3$ and $w=0$, respectively, and thus
\be
\mathcal{P}_s= 
\bc
\frac{4}{9} \Delta_\mathcal{R}^2  \quad \text{(during \rd)}\,, \label{eq:relation between comving and scalar power spectrum} \\
\frac{9}{25} \Delta_\mathcal{R}^2  \quad \text{(during \md)} \,.
\ec 
\ee

Scalar spectral index and running of scalar spectral index are defined as~\cite{Chung:2003iu, Lyth:1998xn, Huang:2006hr} (see also of~\cite{Liddle:2000cg})
\bea
&&n_s = \frac{\td \ln \mathcal{P}_s }{\td \ln k} 
= 1+ 2\eta_V - 6\epsilon_V \,,
\label{eq:ns}
\\
&&\alpha_s \equiv\frac{\td n_s}{\td \ln k}  =16 \epsilon_V \eta_V -24 \epsilon_V^2 - 2\xi_V \,.
\label{eq:alphas}
\eea
The last two terms on the right side of Eq.~\eqref{eq:ns} indicate how much the power spectrum varies from the scale-invariant form. 
The running of the running of scalar spectral index  is defined as~\cite{Chatterjee:2017hru} 
\bea
\beta_s &\equiv&\frac{\td^2 n_s}{\td \ln k^2}   \nonumber \\
&=& 
-192 \epsilon_V^3 + 192 \epsilon_V^2 \eta_V - 32 \epsilon_V \eta_V^2 -24 \epsilon_V \xi_V  + 2\eta_V \xi_V +2 \sigma_V \,.
\eea

The constraints on  $n_s$, $\alpha_s$ and $\beta_s$ from \Planck~observations are mentioned in Table~\ref{Table:PlanckData}.

\subsubsection{Inflection point}
\label{sec:Inflection point}
To find the inflection point of the potential (Eq.~\eqref{eq:inflation potential of model I}), following Eq.~\eqref{eq:condition_for_inflection_point} we need
\ba
\label{eq:condition for inflection point}
 \duP=\dduP=0 \,.
\ea

Solution of Eq.~\eqref{eq:condition for inflection point}
\be\label{eq:inflection point}
\Phi_0 = \frac{3 a}{4 b}  \qquad \text{ for }\, d= \frac{ 8 b^3}{27 a^2} \,.
\ee

Thus, having devised the pathway to find inflection-point, next we move onto understanding how large or small the parameters of the model need to be in order to satisfy the CMB constraints.


\subsubsection{Estimating coefficients from  \Planck~
data}\label{sec:Planck Data}
In this section, we try to fix the coefficients of the potential (Eq.~\eqref{eq:inflation potential of model I}) as a function of $r$. 
Following~\cite{Hotchkiss:2011gz,Chatterjee:2014hna}, we can write
\bea
\begin{pmatrix}
	\Phi_{\rm CMB} & \Phi_{\rm CMB}^2 & \Phi_{\rm CMB}^4\\
	1 & 2\Phi_{\rm CMB} & 4\Phi_{\rm CMB}^3 \\
	0  & 2     & 12\Phi_{\rm CMB}^2 
\end{pmatrix}
\begin{pmatrix}
	a \\
	b\\
	d 
\end{pmatrix}
&&= \begin{pmatrix}
	\uP (\PhC) - V_0\\
	\duP(\PhC)\\
	\dduP(\PhC) 
\end{pmatrix} \,,
\eea
where $d$ is known from Eq.~\eqref{eq:inflection point} and $\uP(\PhC), \duP(\PhC)$ and $\dduP(\PhC)$ can be derived using Eq.~\eqref{eq:epsilonV}, \eqref{eq:etaV},
\eqref{eq:xiV}, \eqref{eq:As}, \eqref{eq:ns}, \eqref{eq:alphas} as
\bea
&&\uP(\PhC) = \frac{3}{2}A_s r \pi^2  \mp^4\, , \label{eq:U}\\
&&\duP(\PhC) = \frac{3}{2}\sqrt{\frac{r}{8}}\left(A_s r \pi^2 \right)  \mp^3\,, \label{eq:U'}\\
&&\dduP(\PhC) = \frac{3}{4}\left(\frac{3r}{8} + n_s -1\right)\left(A_s r \pi^2 \right)  \mp^2\,. \label{eq:U"}
\eea

The observed values of inflation parameters  measured at $\Ph=\PhC$ (at $k_{*}\simeq 0.05 \text{Mpc}^{-1}$) from \Planck~%
data are given in Table~\ref{Table:PlanckData}.
\footnote{  T and E corresponds to temperature and E-mode polarisation of CMB.
}
\begin{table}[ht]
\begin{center}
\caption{ \it \Planck~data of \cmb.} \label{Table:PlanckData}
\begin{tabular}{ |c| c| c|c| }
\hline
$\ln(10^{10} A_s)$ & $3.047\pm 0.014$ & $68\%$, TT,TE,EE+lowE+lensing+BAO & 
\cite{Aghanim:2018eyx}  \\
 \hline
 $n_s$ & $0.9647\pm 0.0043$ & $68\%$, TT,TE,EE+lowE+lensing+BAO & 
 \cite{Aghanim:2018eyx} \\ 
 \hline 
 $\td n_s/\td \ln k$ & $0.0011\pm0.0099$ & $68\%$, TT,TE,EE+lowE+lensing+BAO &  
 \cite{Aghanim:2018eyx}  \\  
 \hline
 $\td^2 n_s/\td \ln k^2$  & $0.009 \pm  0.012$ & $68\%$, TT,TE,EE+lowE+lensing+BAO &  
 \cite{Aghanim:2018eyx} \\
 \hline 
\end{tabular}
\end{center}
\end{table}%
Thus, allowed ranges for 'running of scalar spectral tilt' and 'running of running of spectral tilt' are as follows -
\bea
&& -0.0088 \leq \frac{\td n_s}{\td \ln k} \leq 0.011 \,,\\
&& -0.003 \leq  \frac{\td^2 n_s}{\td \ln k^2} \leq 0.021  \,.
\eea

Constraint on $r$ is
\shila{
\label{r}
r=
 0.014^{+0.010}_{-0.011}\, \text{and}\, <0.036 
 &\quad (95 \%  \,, \text{BK18, \BICEP3, \KeckArray~2020,} \\
 &
 \text{and WMAP and \Planck~CMB polarization)} \no
\\
&
(k_{*}\simeq 0.05 \text{Mpc}^{-1}) \text{(\cite{BICEPKeck:2022mhb,BICEP:2021xfz,Aghanim:2018eyx}, see also~\cite{Campeti:2022vom})} \,. \no
}

\begin{figure}[!h]
\begin{center}
\begin{tabular}[c]{cc}
\begin{subfigure}[c]{0.45\textwidth}
 \includegraphics[width=\linewidth, height=0.6\linewidth ]{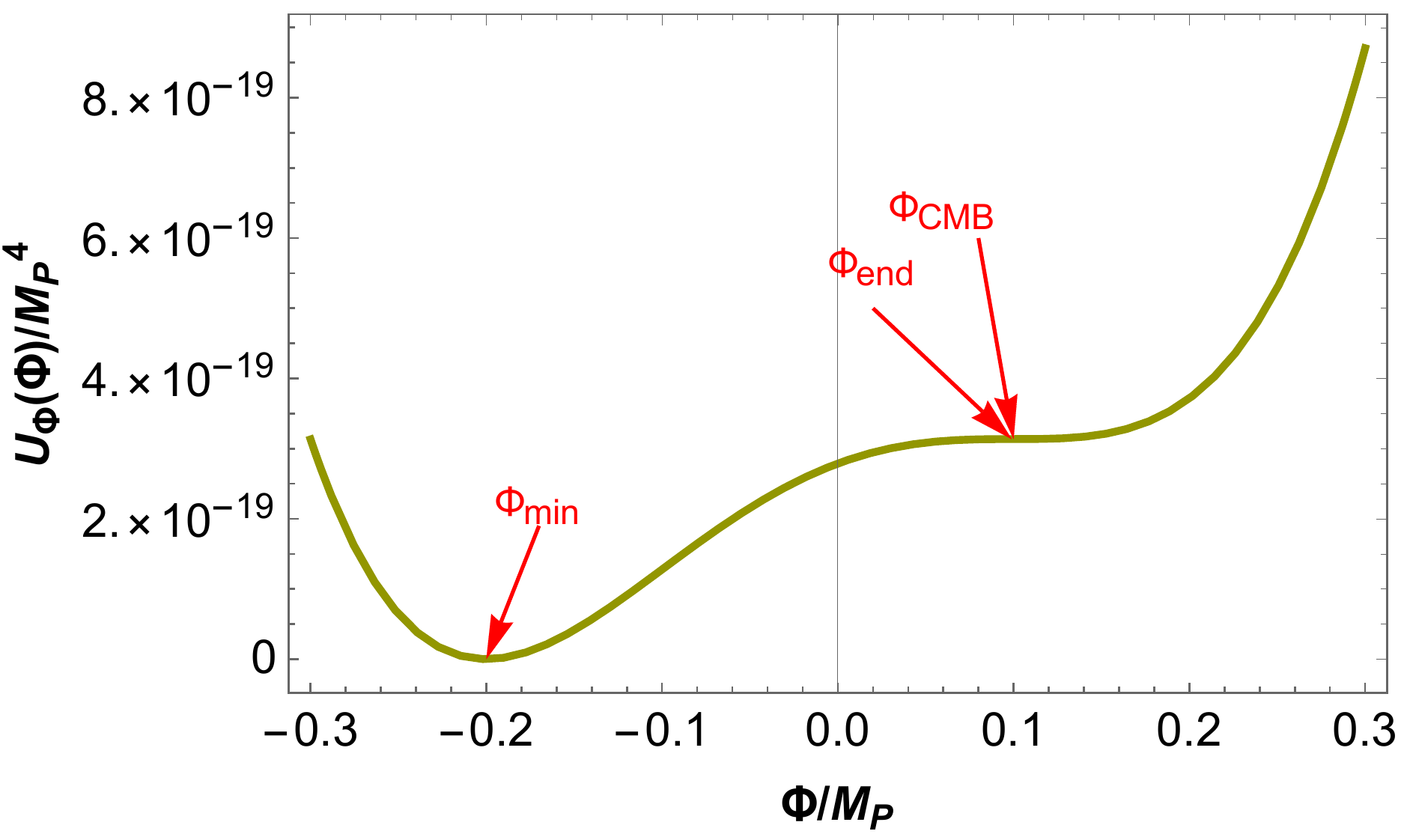}
  \label{fig:mesh1}
\end{subfigure}&%
\hspace{30pt}
\begin{subfigure}[c]{0.45\textwidth}
  \includegraphics[width=\linewidth, height=0.6\linewidth
  ]{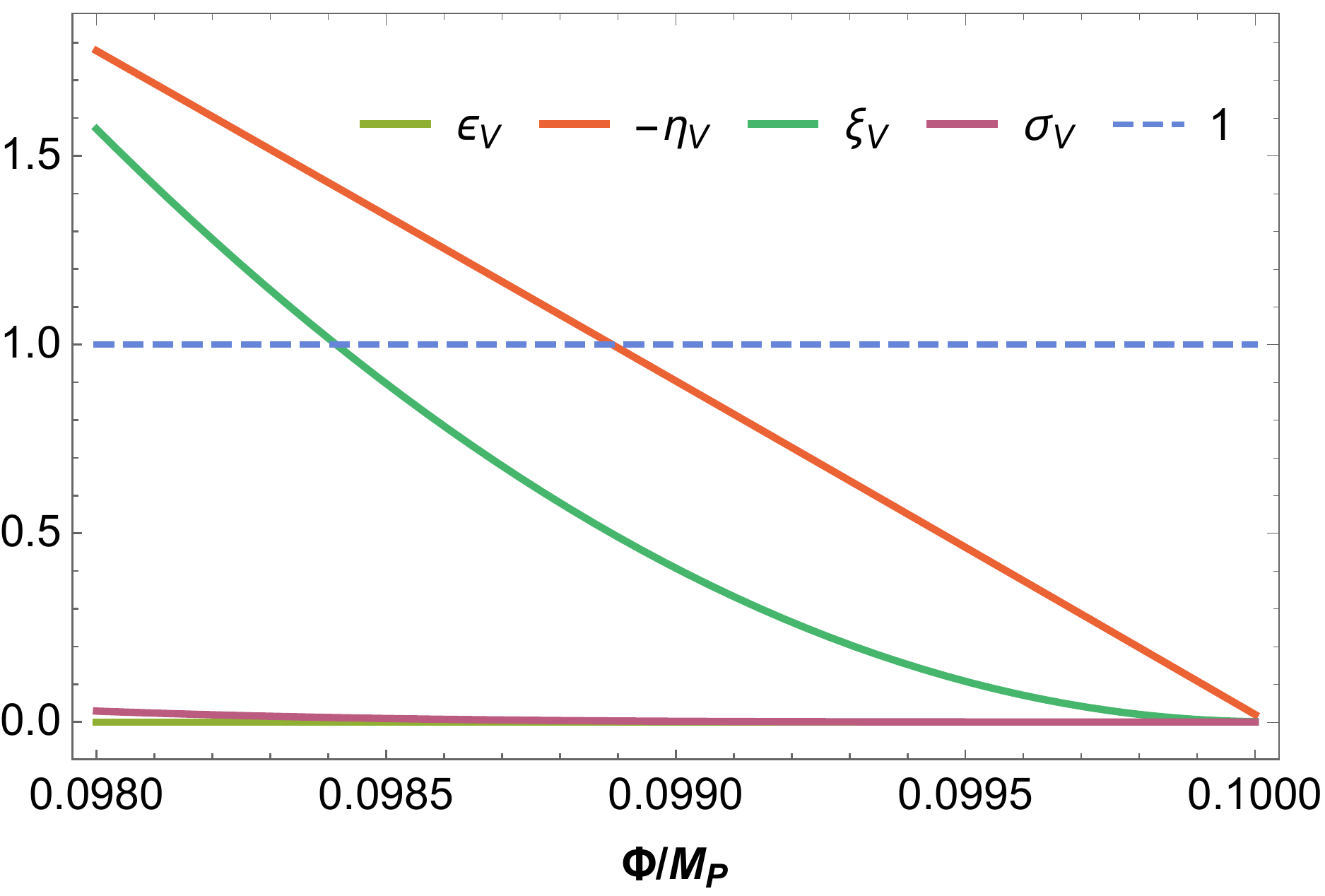}
  \label{fig:sub2}
\end{subfigure}\\
\begin{subfigure}[c]{0.45\textwidth}
 \includegraphics[width=\linewidth, height=0.6\linewidth
 ]{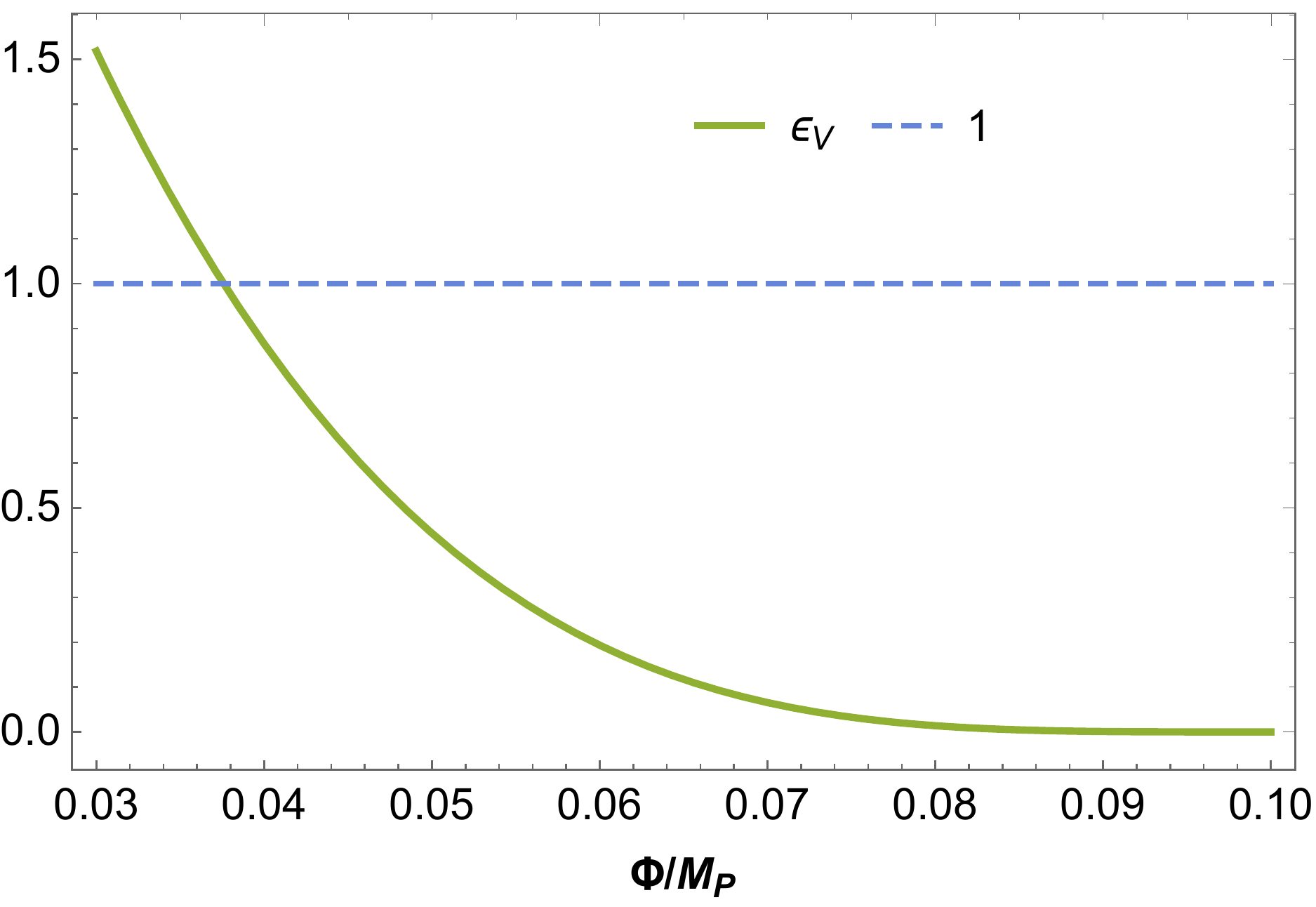}
\end{subfigure}&%
\hspace{30pt}
\begin{subfigure}[c]{.45\textwidth}
  \includegraphics[width=\linewidth, height=0.6\linewidth
  ]{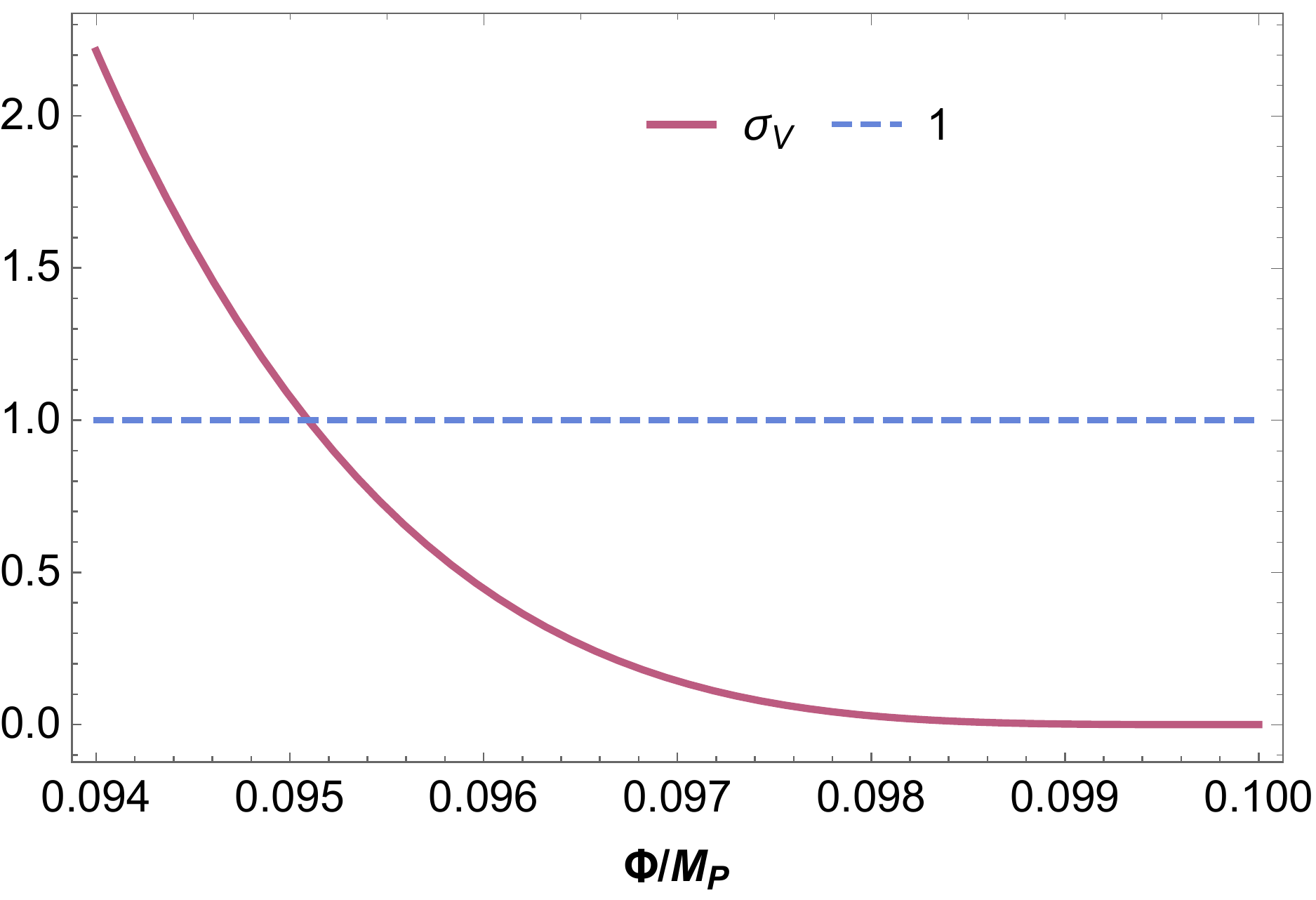}
   \label{fig:sub2}
\end{subfigure}\\
\end{tabular}    

\end{center}
\vspace{-20pt}
\caption{\it \raggedright%
In the top-left panel: normalised inflaton-potential of \mdl I inflation for benchmark value mentioned in Table~\ref{Tab:Model I benchmark values}.  In the top-right panel: The evolution of inflationary slow-roll parameters ($\epsilon_V,-\eta_V, \xi_V,\sigma_V$) as a function of $\Ph/\mp$; second row - left panel: $\epsilon_V$, and second row – right panel: $\sigma_V$ of \mdl I slow roll inflation as a function of $\Ph/\mp$ are shown separately for benchmark values mentioned in Table~\ref{Tab:Model I benchmark values}. 
The dashed line is for $1$. As soon as $\lt|\eta_V\rt|\sim 1$, the slow roll inflation ends. From these figures, it is clear that $\lt|\epsilon_V \rt|<\lt| \sigma_V\rt|< \lt|\xi_V \rt|<\lt|\eta_V\rt|$ during the slow-roll regime.
}
\label{fig:potential_plot_linear_term_inflation}
\end{figure}

Using these, we can find the coefficients of the potential. However, it is sufficient to design the potential in a way such that $\PhC$ is adjacent to $\Ph_0$~\cite{Garcia-Bellido:2017mdw} (see also~\cite{Okada:2016ssd}). In order to implement this, let us revamp the potential (Eq.~\eqref{eq:inflation potential of model I}) as
\ba\label{eq:modified potential of model I}
\UP = V_0 + A \, \Ph - B \, \Ph^2 + d \, \Ph^4 \,,
\ea 
with $A= a (1-\b^I_1), B= b(1-\b^I_2)$ (where $\b^I_1, \b^I_2$ are dimensionless) such that in the limit $\b^I_1,\b^I_2 \to 0$, the slope of the potential vanishes at $\PhZ$. Using these arrangements, 
we have shown the benchmark value for this potential in Table~\ref{Tab:Model I benchmark values} and using this value, the evolution of the potential and slow roll parameters with $\Ph$ are illustrated in Fig.~\ref{fig:potential_plot_linear_term_inflation}. At $\Ph=\PhC$,  $\epsilon_V,\lt|\eta_V\rt|, \xi_V,\sigma_V <<1$, and at $\Ph=\Ph_{\rm end}$, $\lt|\eta_V\rt| \simeq  1$ (from Fig.~\ref{fig:potential_plot_linear_term_inflation} it is clear that $\sigma_V,\xi_V,\epsilon_V<\lt|\eta_V\rt|$) leads to the ending of slow roll phase.


\begin{table}[H]
    \centering
        \caption{ \it Benchmark values for linear term potential (\mdl I) ($\Ph_{\rm min}$ is the minimum of potential in Eq.~\eqref{eq:modified potential of model I}) }
    \label{Tab:Model I benchmark values}
\vspace{-20pt}
\begin{center}
\begin{tabular}{ |c| c | c|c|c| c| }
\hline 
$V_0/\mp^4$ & $a/\mp^3$ & $b/\mp^2$ & $d$ & $\b^I_1$ & $\b^I_2$\\
 \hline 
 $2.788\times 10^{-19}$ &  $9.29\times 10^{-19}$   & $6.966\times 10^{-18}$  & $1.16\times 10^{-16}$   & $6 \times 10^{-7}$ & $6 \times 10^{-7}$ \\
 \hline 
\end{tabular}

\vspace{0.5pt}
\begin{tabular}{ |c| c |c|c| c |}
\hline
$\Ph_{\rm CMB}/\mp$ & $\Ph_{\rm end}/\mp$ & $\Ph_{\rm min}/\mp$         & $\PhZ/\mp$\\
\hline 
 $0.1$ & $0.098889 $ & $-0.200045$    &      $0.100022$ \\
\hline
\end{tabular}

\vspace{0.5pt}
\begin{tabular}{ |c| c |c|c| c |c|c|}
\hline
$r$ & $n_s$ & $A_s$  & e-folding & $\alpha_s$ & $\beta_s$ &  $\Delta {\cal R}^2$ \\
\hline 
$9.87606\times 10^{-12}$ & $0.960249$ & $2.10521\times 10^{-9} $ & $53.75$ & $-1.97\times 10^{-3}$ & $ -3.92\times10^{-5}$ & $2.14496\times 10^{-9}$ \\
 \hline 
\end{tabular}
\end{center}
\end{table}


\subsection{Inflection-point achieved with Sextic Term}
\label{sec:Inflection-point achieved with Sextic Term}

Here, we discuss the inflationary scenario for the potential of \mdl II. Similar to the \mdl I, we are again assuming that there is no coupling between the inflaton $\vp$ and the gravity.  
Thus, the Lagrangian density of this case is similar to Eq.~\eqref{Eq:Lagrangian density-ModI}, except for three terms, as indicated below-%
\shila{
\label{Eq:Lagrangian density Model-II}
\mathcal{L}_{II} &= \frac{\mp^2}{2} \mathcal{R}
+\mathcal{L}_{INF} (\vp)
+ {\cal L}_{KE, \chi} 
  - U_\chi(\chi) 
  + {\cal L}_{KE, H} 
  - U_H(H)
 + \mathcal{L}_{reh, II}%
 + \mathcal{L}_{SM} \,,
}
%
where Lagrangian density responsible for inflation is
\ba 
\mathcal{L}_{INF} (\vp)= \frac{1}{2}\partial_\mu \vp\partial^\mu\vp - U_\vp(\vp) \,,
\ea 
and the 
potential for inflaton is given by~\cite{Bhaumik:2019tvl}:
\ba\label{eq:inflation potential of model II}
U_\vp(\vp)  = 
p \, \vp^2 - q \, \vp^4 + \tw \, \vp^6 \,.
\ea 
We are assuming that $p, q$, and $\tw$ are all reals and $\tw>0$. These coefficients have dimensions of $
\mp^2, \mp^0$, and $\mp^{-2}$, respectively. Such sextic terms are motivated may have several origins ranging from EFT considerations~\cite{Penco:2020kvy} to SUSY-based considerations~\cite{Lyth:1995ka,Kawasaki:2009hp,Chatterjee:2017hru,Binetruy:1986ss,Lazarides:1985ja,Lazarides:1986rt,Lazarides:1992gg,Jokinen:2004bp} and several more. Again here, we do not go into details of the origins and instead treat this as phenomenological analysis which will remain to be applicable for all such models. Now, the Lagrangian density for the reheating is
\ba\label{eq:reheating lagrangian for modelII}
\mathcal{L}_{reh, II} = - \yc \vp \cc - \lO \vp \HH - \lT \vp^2 \HH \,.
\ea 
This potential of Eq.~\eqref{eq:inflation potential of model II} has an inflection point at
\ba
\vpZ= \frac{\sqrt{q}}{\sqrt{3\,  \tw}} \qquad \text{ for } \, p= \frac{ q^2}{3 \, \tw} \,.
\ea 


\begin{figure}[htp]
\centering
\begin{subfigure}{0.45\textwidth}
  \centering
 \includegraphics[width=\linewidth]{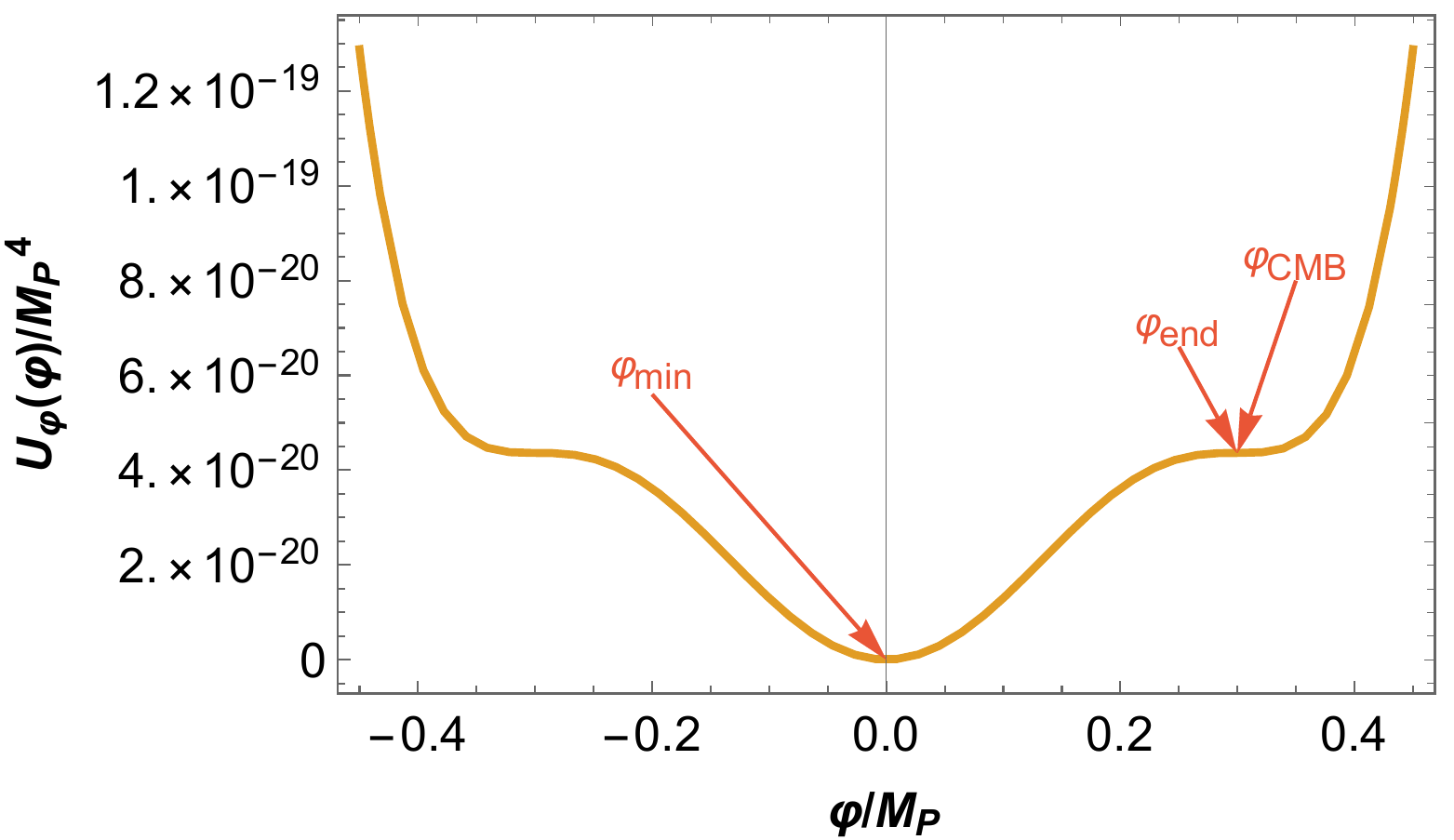}
  \label{fig:mesh1}
\end{subfigure}%
\hspace{30pt}
\begin{subfigure}{.45\textwidth}
  \centering
  \includegraphics[width=\linewidth]{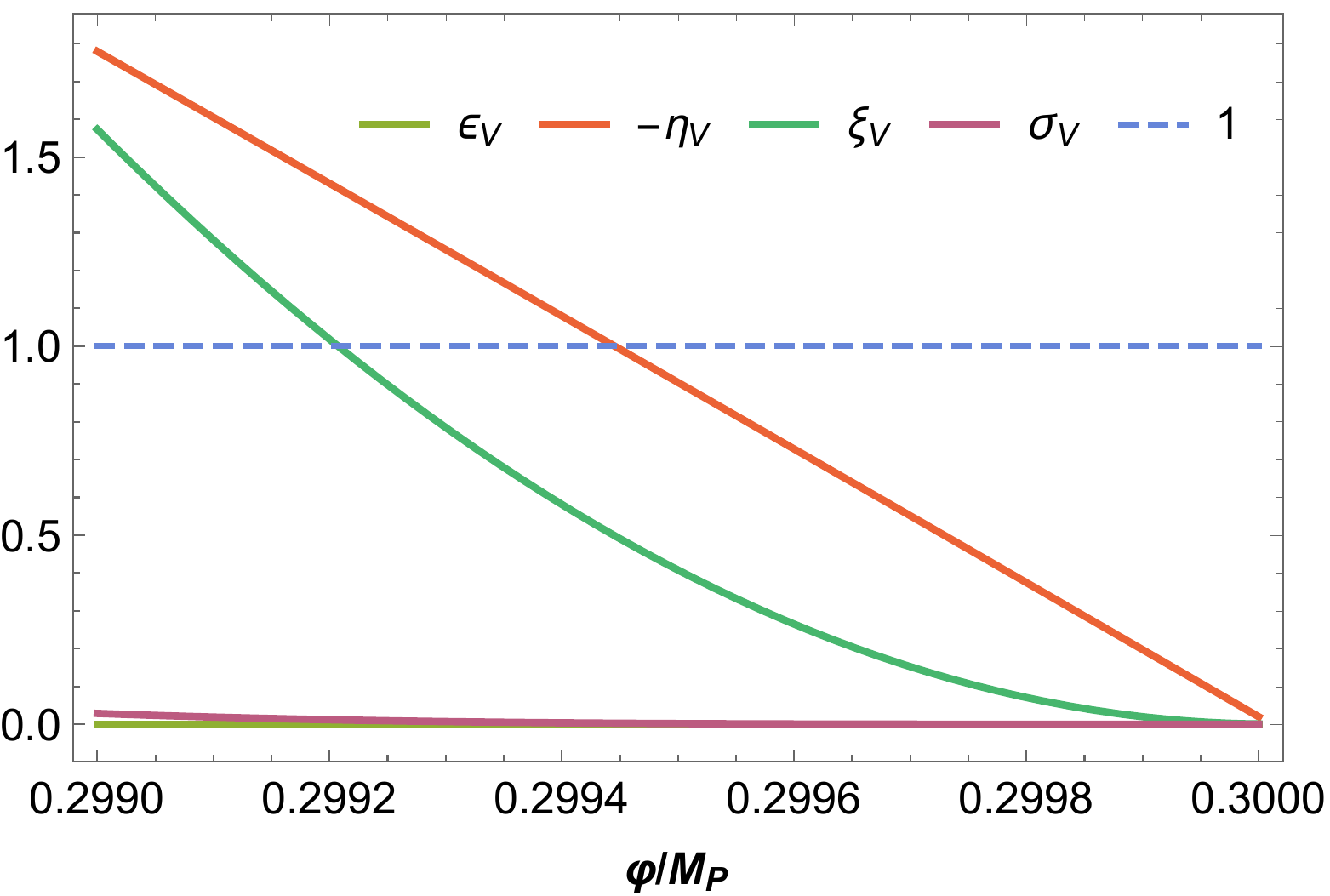}
  \label{fig:sub2}
\end{subfigure}
\begin{subfigure}{0.45\textwidth}
  \centering
 \includegraphics[width=\linewidth]{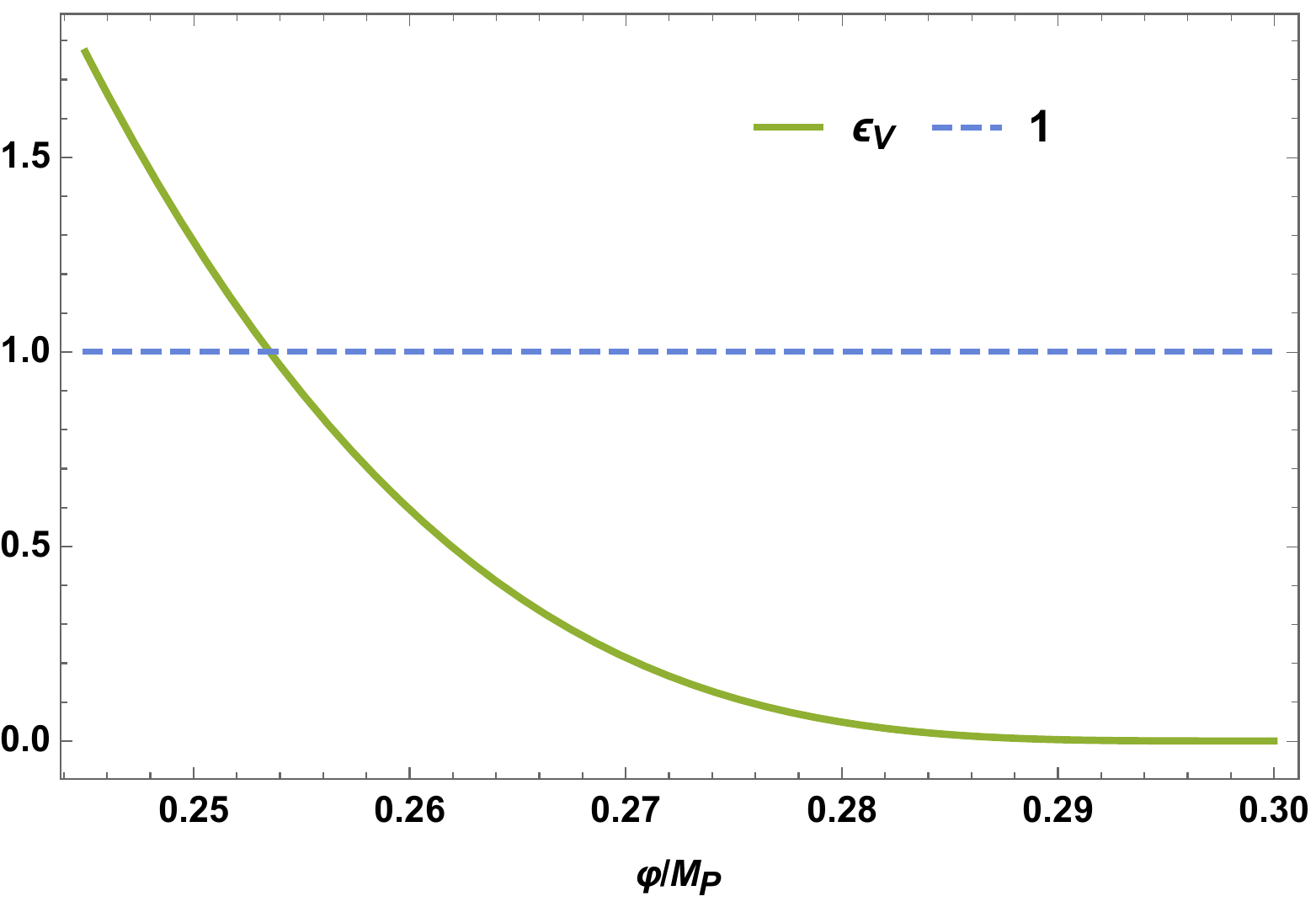}
   \label{fig:mesh1}
\end{subfigure}%
\hspace{30pt}
\begin{subfigure}{.45\textwidth}
  \centering
  \includegraphics[width=\linewidth]{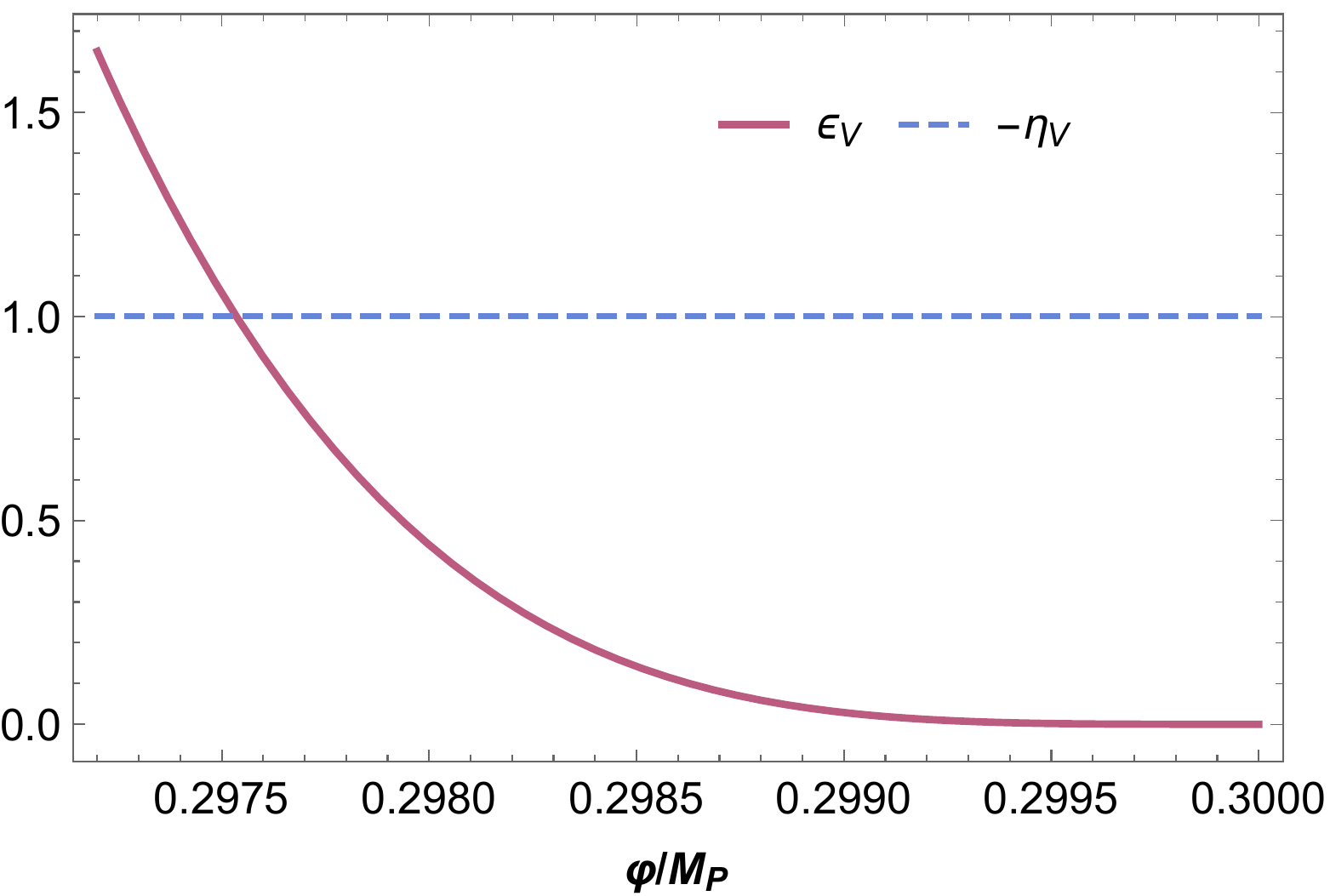}
   \label{fig:sub2}
\end{subfigure}
\vspace{-30pt}
\caption{\it \raggedright In the top-left panel, normalised inflaton-potential of \mdl II as a function of '$\vp/\mp$' for benchmark value mentioned in Table~\ref{Tab:Model II benchmark values}. In the top-right panel, absolute values of four slow roll parameters ($\epsilon_V, -\eta_V,\xi_V,\sigma_V $) are plotted with $\vp/\mp$ as variable. Left and righ panel of the second row displays $\epsilon_V$ and $\sigma_V$ respectively against $\vp/\mp$ for benchmark values mentioned in Table~\ref{Tab:Model II benchmark values}. 
The dashed line indicates $1$. 
These graphs show that $\lt|\epsilon_V \rt|<\lt| \sigma_V\rt|< \lt|\xi_V \rt|<\lt|\eta_V\rt|<1$ during the slow-roll inflation, similar to what we have obtained in \mdl I. 
}
\label{fig:potential_plot_sextic_inflation}
\end{figure}

Following the definition from Section~\ref{sec: slow roll parameters}, the inflationary slow roll parameters are 
\bea
&&\epsilon_V= \mp^2\frac{2 \left(p \varphi -2 q \varphi ^3+3 \tw \varphi ^5\right)^2}{\left(p \varphi ^2-q \varphi ^4+\tw \varphi ^6\right){}^2} \,, \\
&&\eta_V= \mp^2\frac{2 \left(p-6 q \varphi ^2+15 \tw \varphi ^4\right)}{p \varphi ^2-q \varphi ^4+\tw \varphi^6}  \,, \\
&&\xi_V =\mp^4\frac{48 \varphi ^2 \left(-q+5 \tw \varphi ^2\right) \left(p-2 q \varphi ^2+3 \tw \varphi  ^4\right)}{\left(p \varphi ^2-q \varphi ^4+\tw \varphi ^6\right){}^2}  \,,\\
&&\sigma_V=\mp^6\frac{96 \left(-q+15 \tw \varphi ^2\right) \left(p \varphi -2 q \varphi ^3+3 \tw \varphi
   ^5\right)^2}{\left(p \varphi ^2-q \varphi ^4+\tw \varphi ^6\right){}^3}   \,.
\eea 

The slow roll inflation will continue as long as the kinetic energy of the inflaton is negligible compared to the potential energy $\uvp(\vp)$ and as soon as $\lt|\epsilon_V\rt|,\lt|\eta_V\rt|, \lt|\xi_V\rt|, \lt|\sigma_V\rt| \sim 1$ at $\vp=\vp_{\rm end}$, the inflaton aborts the slow-roll phase.
Following Section~\ref{sec:Inflection-point achieved with Linear term}, we can also define $N_{\rm CMB}, A_s, n_s, \a_s$, and $\b_s$ for the potential $\uvp(\vp)$. Similarly, using the approach of Section~\ref{sec:Planck Data}, we redefine the potential as 
%
\ba\label{eq:modified potential of model II}
U_\vp(\vp)  = 
p \, \vp^2 - Q \, \vp^4 + \text{W} \, \vp^6 \,,
\ea 
such that $Q= q(1-\b^{II}_1)$ and $\text{W}=w (1-\b^{II}_2)$ and $\b^{II}_1$, $\b^{II}_2$ have zero mass dimension.
Then, we can estimate $p,q$ and $\tw$, and the values are mentioned in Table~\ref{Tab:Model II benchmark values}. For this value, the variation of $\uvp(\vp)$ of Eq.~\eqref{eq:modified potential of model II} and $\epsilon_V,\lt|\eta_V\rt|, \xi_V,\sigma_V $ as a function of $\vp$ is shown in Fig.~\ref{fig:potential_plot_sextic_inflation}.
The slow roll inflationary phase ends at $\vp_{\rm end}$ when  $\lt|\eta_V\rt|\simeq 1$  (because for \mdl II $\epsilon_V<\lt|\eta_V\rt|$).


\begin{table}[H]
    \centering
        \caption{ \it Benchmark values for sextic potential ($\vp_{\rm min}$ is the minimum of potential Eq.~\eqref{eq:modified potential of model II})}
    \label{Tab:Model II benchmark values}
\vspace{-20pt}
\begin{center}
\begin{tabular}{ | c |c|c| c |c|}
\hline 
 $p/\mp^2$ & $q$ & $\tw \mp^2$ & $\b^{II}_1$     & $\b^{II}_2$\\
 \hline 
  $1.45\times 10^{-18}$   & $1.62\times 10^{-17}$  & $5.98\times 10^{-17}$   &  $1.53\times 10^{-8}$   & $1.53\times 10^{-8}$ \\
 \hline 
\end{tabular}

\vspace{0.5pt}
\begin{tabular}{ |c| c |c|c| c |}
\hline
$\vp_{\rm CMB}/\mp$ & $\vp_{\rm end}/\mp$ & $\vp_{\rm min}/\mp$         & $\vpZ/\mp$\\
\hline 
$0.3$ & $0.299444$ & $ 0 $     &  $0.300011$\\
\hline
\end{tabular}

\vspace{0.5pt}
\begin{tabular}{ |c| c |c|c| c |c|c|  }
\hline
$r$ & $n_s$ & $A_s$  & e-folding & $\alpha_s$ & $\beta_s$ &  $\Delta {\cal R}^2$       \\
\hline 
$1.4 \times 10^{-12}$ & $0.96001$ & $2.10521\times 10^{-9} $ & $60.247$ & $-1.487\times 10^{-3}$ & $ -2.972\times 10^{-5}$ & {$2.10521\times 10^{-9} $}  \\
 \hline 
\end{tabular}
\end{center}
\end{table}

\section{Stability analysis}
\label{Sec:Stability analysis}
In this section, we try to find the upper bound of $\yc$ and $\lO$ so that ${\cal L}_{reh,I}$ and ${\cal L}_{reh,II}$ do not affect the inflationary scenario set forth in Section~\ref{sec:Inflection-point achieved with Linear term} and in Section~\ref{sec:Inflection-point achieved with Sextic Term}, respectively. 
The Coleman–Weinberg (CW) correction
to the inflaton-potential, which is originated as radiative correction at 1-loop order, is given by~\cite{Drees:2021wgd} -
\begin{eqnarray}
V_{\rm CW}
=\sum_{j
} \frac{n_j}{64\pi^2} (-1)^{2s_j}\widetilde{m}_j^4
\left[ \ln\left( \frac{\widetilde{m}_j^2 
}{\mu^2} \right) - c_j \right]  \,.
\end{eqnarray}

Here $\widetilde{m}_j$ is inflaton dependent mass of the component $j$. 
The sum over $j$ is actually over three fields - inflaton, $\c$, and $H$. The counter term $c_j=\frac{3}{2}$%
; $n_{H,\c}=4$, 
$s_H =0$
, and $s_\c=1/2$. $n_j$ and $s_j$ for inflaton are $1$ and $0$, respectively. 
$\mu$ is the renormalization scale, which is taken $\sim
$ $\PhZ$ \fmIt or $\vpZ$ \fmIIt.
The first and second derivative of the CW term w.r.t. inflaton
\shila{
	 & V_{\rm CW}^\prime = \sum\limits_{j 
	 }\frac{n_j}{32 \pi^2}  (-1)^{2 s_j} 
	\widetilde{m}_j^2 \, \lt(\widetilde{m}_j^{2}\rt)^{\prime}
	\left[ \ln \left(\frac{\widetilde{m}_j^2}{\mu^2} \right)
	- 1 \right]\,,  \label{eq:first derivative test}\\
	& V_{\rm CW}^{\prime\prime} =  \sum\limits_{j
	} \frac{n_j}{32 \pi^2} (-1)^{2 s_j} 
	\left\{ \left[ \left(\lt(\widetilde{m}_j^{2}\rt)^{\prime} \right)^2
	+ \widetilde{m}_j^2 \lt(\widetilde{m}_j^{2}\rt)^{\prime\prime} \right]
	\ln \left(\frac{\widetilde{m}_j^2}{\mu^2} \right)
	- \widetilde{m}_j^2 \lt(\widetilde{m}_j^{2}\rt)^{\prime\prime}  \right\}\,.   \label{eq:second derivative test}
}
%
In the next two subsections, we discuss the stability 
relative to the couplings $\yc$ and $\lO$ for the two inflation-potentials
~(Eq.~\eqref{eq:modified potential of model I}) and Eq.~\eqref{eq:modified potential of model II}) we have considered.

\subsection{Stability analysis for linear term inflation}
\label{sec:Stability analysis for linear term inflation}

For the inflationary scenario of Section~\ref{sec:Inflection-point achieved with Linear term}, the field-depended mass of the fermionic field and Higgs field are respectively 
\ba
& \widetilde{m}_{\c}^2 (\Ph) = \lt( m_\c + y_\c \Ph \rt)^2  \,,\\
& \widetilde{m}_{H}^2 (\Ph) = m_H^2 + \lO \Ph \,.
\ea 

For the stability of the inflation-potential, the terms of the order of $\lO^2$ and $\yc^2$  on the right-hand side in Eq.~\eqref{eq:first derivative test} and Eq.~\eqref{eq:second derivative test} should be less than corresponding tree level terms from 
Eq.~\eqref{eq:modified potential of model I} -
\shila{
& V_{\rm tree}^\prime (\PhZ) \equiv \duP(\PhZ) = \frac{32 b^3 \PhZ^3}{27 a^2}+a (1-\beta )-2 b (1-\beta ) \PhZ\,,
\label{eq:tree-level-first-derivative-modelI} \\
& V_{\rm tree}^{\prime\prime} (\PhZ) \equiv \dduP(\PhZ) =  \frac{32 b^3 \PhZ^2}{9 a^2}-2 b (1-\beta )
\,,\label{eq:tree-level-second-derivative-modelI}
} 
where $\b^I_1 = \b^I_2=\b^I$ (for our chosen benchmark value $\b^I_1 = \b^I_2$).

The first derivative (Eq.~\eqref{eq:first derivative test}) and second derivative (Eq.~\eqref{eq:second derivative test}) of CW term for Higgs field are 
\shila{
& \lt| V_{{\rm CW},H}^\prime  
\rt|=\frac{\lO^2 \Phi }{8 \pi ^2}  \left(\ln
\left(\frac{\lO \Phi }{\Ph_0^2}\right)-1\right)  \, , \label{eq:stability-H-first derivative}\\
& \lt|V_{{\rm CW},H}^{\prime \prime}
\rt|= \frac{\lO^2 }{8 \pi ^2} \ln
\left(\frac{\lO \Phi }{\Ph_0^2}\right) \, .\label{eq:upper limit of l12-coupling model I}
}

\begin{figure}[htp]
\centering
\begin{subfigure}{0.45\textwidth}
  \centering
 \includegraphics[width=\linewidth]{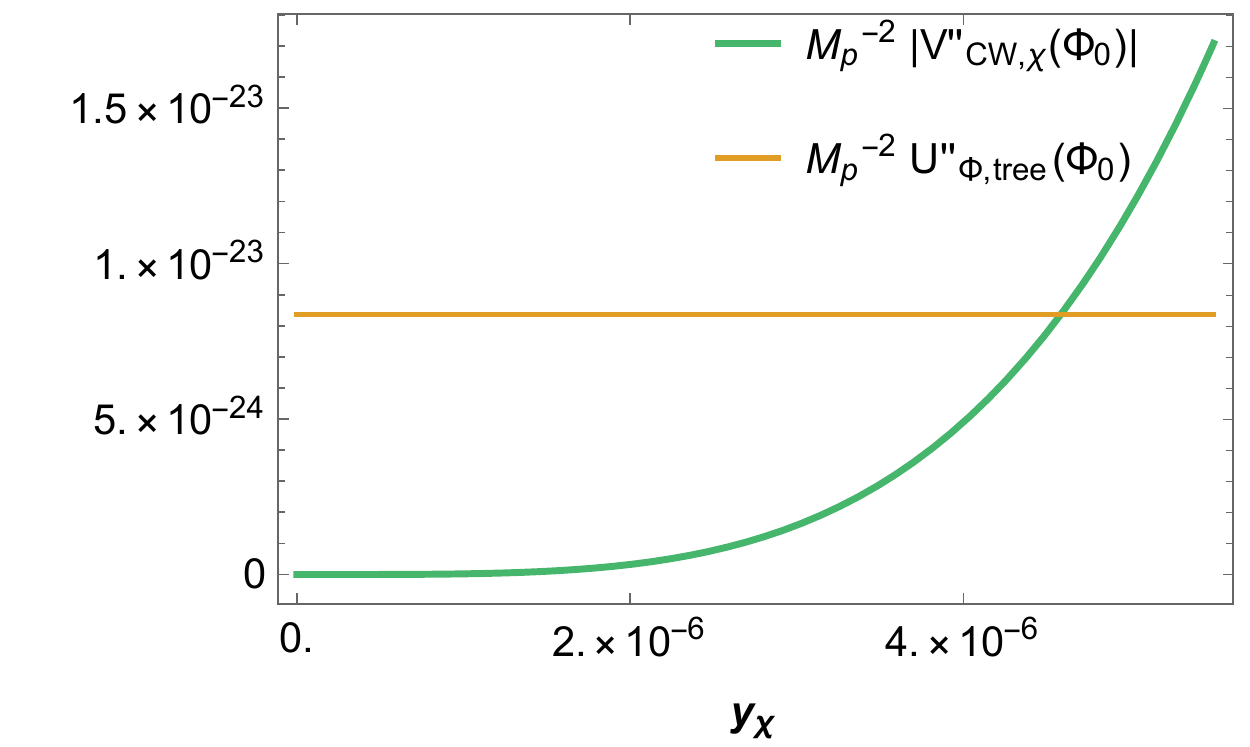}
\end{subfigure}%
\hspace{30pt}
\begin{subfigure}{.45\textwidth}
  \centering
  \includegraphics[width=\linewidth]{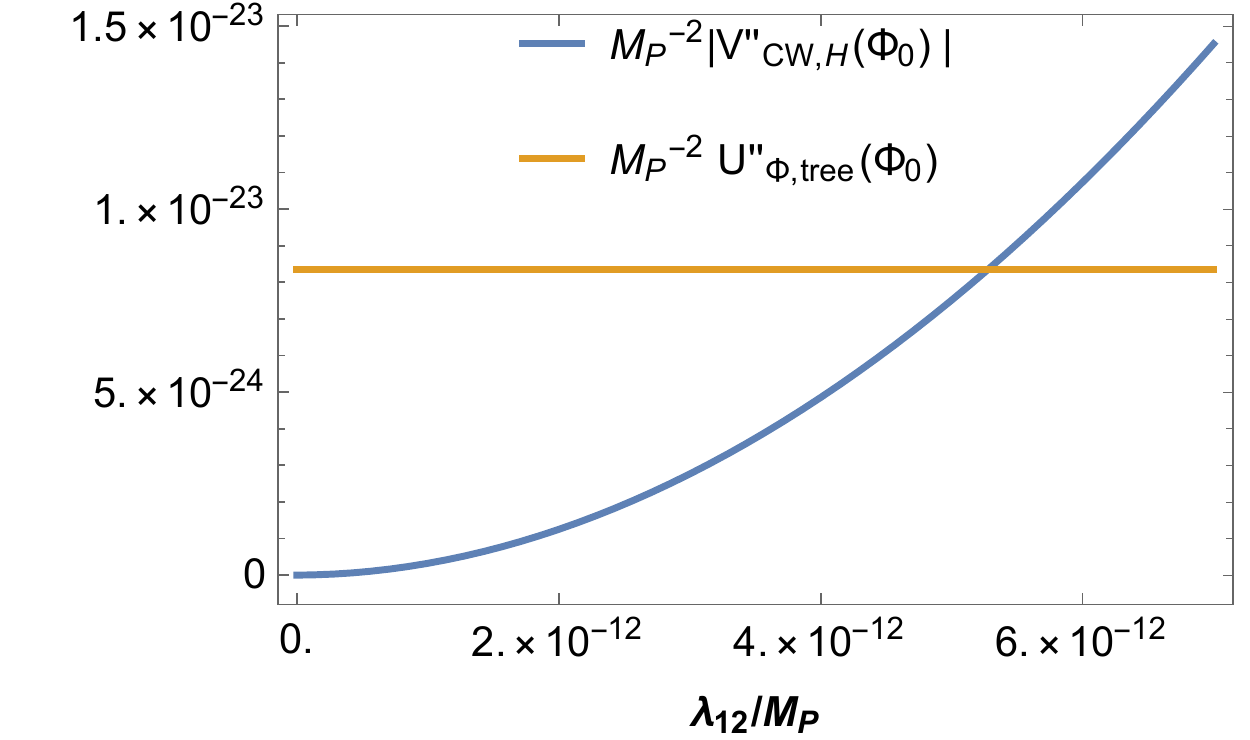}
\end{subfigure}
\vspace{-10pt}
\caption{\it  \raggedright
Allowed range for $\yc$ and $\lO$ for \mdl I inflation from stability. 
The yellow colored line corresponds to tree level potential of $\Ph$. The green and blue colored line represents the CW correction due to $\chi$ and $H$. 
}
\label{fig:stability_analysis_plot_for_model_I}
\end{figure}

The upper bound of the value of $\lO$  at $\Ph\sim\PhZ$ comes from $\lt|V_{{\rm CW},H}^{\prime \prime}\rt| 
< V_{\rm tree}^{\prime \prime} (\PhZ)$, and it is shown on the right panel of Fig.~\ref{fig:stability_analysis_plot_for_model_I}. Thus, $\lO/\mp$ must be $<5.283\times 10^{-12}$. 

Similarly,  for coupling to $\chi$,
\shila{
&\lt| V_{{\rm CW},\c}^\prime  
\rt|  = \frac{\Phi ^3 \yc^4 }{4 \pi ^2} \left( 1 - \ln
\left(\frac{\Phi ^2 \yc^2}{\Ph_0^2}\right)\right) \, ,\\
& \lt|V_{{\rm CW},\c}^{\prime \prime} 
\rt|= \frac{1}{8 \pi ^2} \lt(6 \Phi ^2 \yc^4  \ln
\left(\frac{\Phi ^2 \yc^2}{\Ph_0^2}\right)-2 \Phi ^2 \yc^4 \rt)\, .
\label{eq:upper limit of yc model I}
}
The upper bound on $\yc$ around $\Ph\sim\PhZ$ can be obtained from $\lt|V_{{\rm CW},\c}^{\prime \prime} \rt|<V_{\rm tree}^{\prime\prime} (\PhZ)$ which is exhibited in the left panel of Fig.~\ref{fig:stability_analysis_plot_for_model_I}, and it gives $\yc < 4.578\times 10^{-6}$.

\subsection{Stability analysis for sextic inflation}

In this model, inflaton is $\vp$. Accordingly, the field-depended mass of the fermionic field and Higgs field are respectively 
\ba
& \widetilde{m}_{\c}^2 (\vp) = \lt( m_\c + y_\c \vp \rt)^2 \,,\\
& \widetilde{m}_{H}^2 (\vp) = m_H^2 + \lO \vp \,.
\ea 

\begin{figure}[htp]
\centering
\begin{subfigure}{0.45\textwidth}
  \centering
 \includegraphics[width=\linewidth]{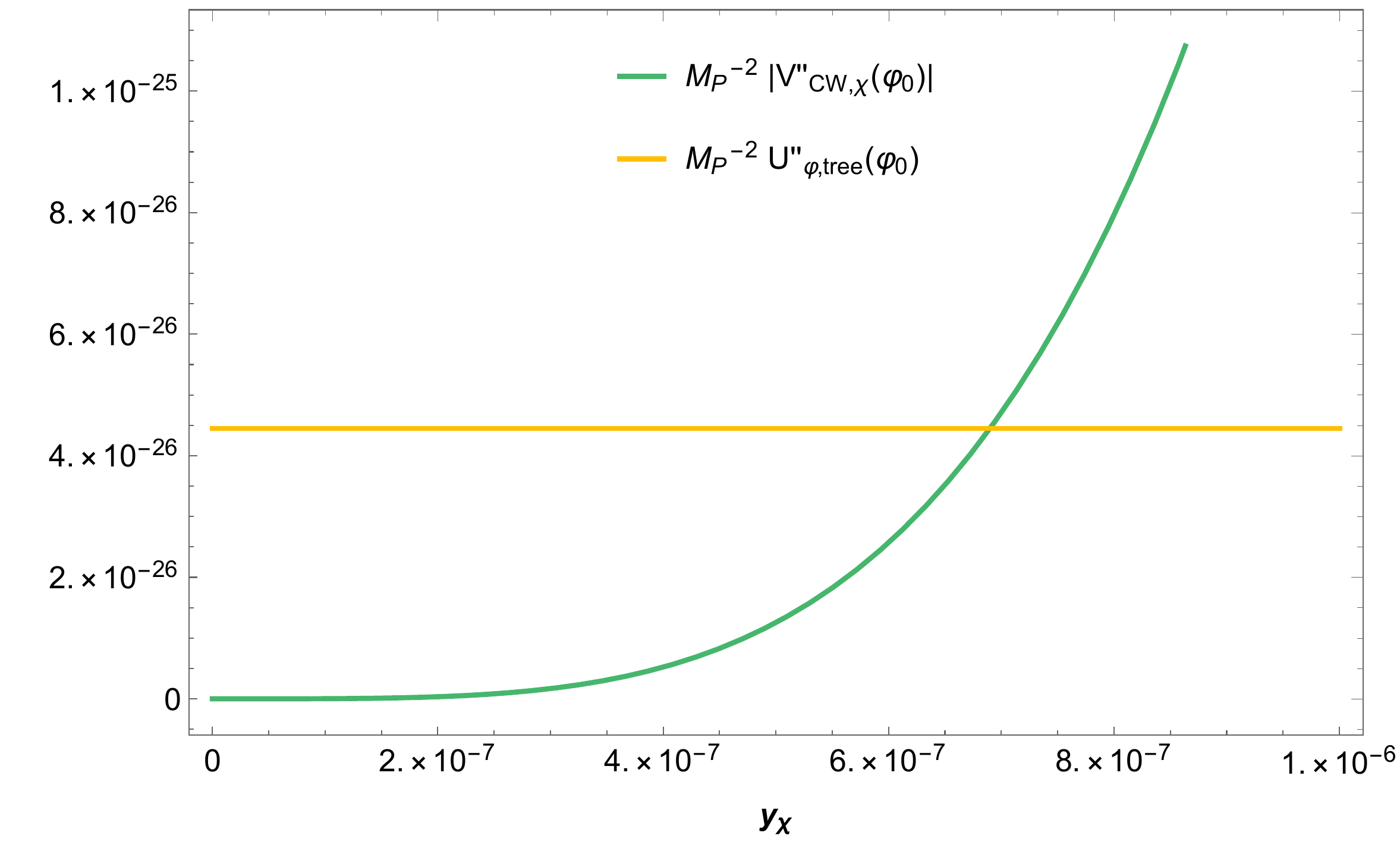}
\end{subfigure}%
\hspace{30pt}
\begin{subfigure}{.45\textwidth}
  \centering
  \includegraphics[width=\linewidth]{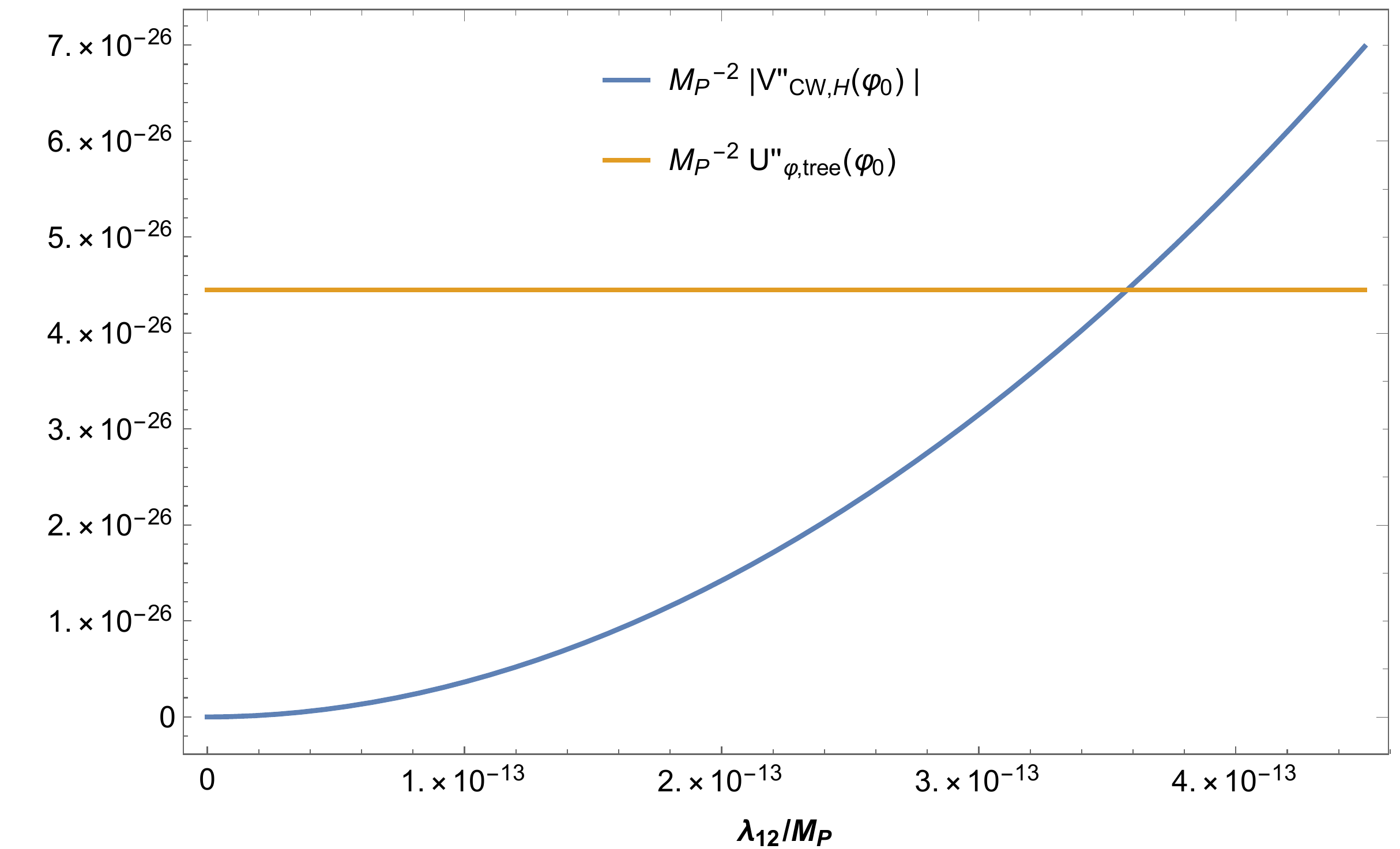}
\end{subfigure}
\vspace{-10pt}
\caption{\it  \raggedright
Allowed range for $\yc$ and $\lO$ for \mdl II inflation from stability analysis of the inflaton-potential. 
Green and blue colored line are due to CW correction for $\c$ and $H$, and they are compared with 
tree level potential of $\vp$ (yellow colored horizontal line). 
}
\label{fig:stability_analysis_plot_for_model_II}
\end{figure}

From Eq.~\eqref{eq:modified potential of model II}
\shila{
& V_{\rm tree}^\prime (\vpZ) \equiv \duvp(\vpZ) = \frac{2 q^2 \vpZ}{3 \tw}-4 (1-\beta^{II} ) q \vpZ^3+6 (1-\beta^{II} ) \tw \vpZ^5 \,,\\
& V_{\rm tree}^{\prime\prime} (\vpZ) \equiv \dduvp(\vpZ) = \frac{2 q^2}{3 \tw}-12 (1-\beta^{II} ) q \vpZ^2+30 (1-\beta^{II} ) \tw \vpZ^4 \,,
}

where $\b^{II}_1=\b^{II}_2=\b^{II}$ (because we have chosen  $\b^{II}_1=\b^{II}_2$ in our benchmark value).
Following the steps similar to the ones mentioned in Section~\ref{sec:Stability analysis for linear term inflation}, for $\lO$ (Eq.~\eqref{eq:first derivative test}) and (Eq.~\eqref{eq:second derivative test}) results in
\shila{
&\lt| V_{{\rm CW},H}^\prime 
\rt|=\frac{\lO^2 \vp }{8 \pi ^2}  \left(\ln
\left(\frac{\lO \vp }{\vpZ^2}\right)-1\right)  \, , \label{eq:upper limit of l12-coupling model II}\\
& \lt|V_{{\rm CW},H}^{\prime \prime}  
\rt|= \frac{\lO^2 }{8 \pi ^2} \ln
\left(\frac{\lO \vp }{\vpZ^2}\right) \, ,
}
and for $\yc$
\shila{
& \lt|V_{{\rm CW},\c}^\prime 
\rt|= \frac{\vp ^3 \yc^4 }{4 \pi ^2} \left( 1 -\ln
\left(\frac{\vp ^2 \yc^2}{\vpZ^2}\right)\right) \, , \\
& \lt| V_{{\rm CW},\c}^{\prime \prime}
\rt|= \frac{1}{8 \pi ^2} \lt(6 \vp ^2 \yc^4  \ln
\left(\frac{\vp ^2 \yc^2}{\vpZ^2}\right)-2 \vp ^2 \yc^4 \rt). \label{eq:upper limit of yc model II}
}
In this inflationary case, upper bound on $\lO$ and $\yc$  around $\vp\sim\vpZ$ comes from $\lt|V_{{\rm CW},H}^{\prime\prime} \rt|< V_{\rm tree}^{\prime\prime} (\vpZ)$, and $\lt|V_{{\rm CW},\c}^{\prime \prime} \rt|<V_{\rm tree}^{\prime\prime} (\vpZ)$, respectively. These have been shown in Fig.~\ref{fig:stability_analysis_plot_for_model_II}.  The upper bounds are $\yc < 6.9 \times 10^{-7}$, and $\lO/\mp< 3.58 \times 10^{-13}$. And lower bound of $\lO$ (it is not shown in Fig.~\ref{fig:stability_analysis_plot_for_model_II}) appears in Eq.~\eqref{eq:lower limit of coupling}.

\section{Reheating and Dark Matter}
\label{Sec:Reheating}
At the end of slow roll phase, the inflaton  
initiates an epoch of coherent oscillation around  
the minimum of the inflationary potential. In contrast to slow roll regime, this epoch is highly adiabatic. Furthermore,
we assume that this oscillating field when averaged over a number of coherent oscillations~\cite{Kolb:2003ke},
~behaves as a non-relativistic fluid but without any pressure~(\cite{Mukhanov:2005sc}, see also~\cite{Garcia:2020eof,Garcia:2020wiy}).
%
Due to the Hubble expansion of the universe, the energy of the oscillating inflaton decreases, and as a result, the amplitude of oscillation 
diminishes~\cite{Allahverdi:2010xz}). 
Together with that, the inflaton-energy is also converted possibly to relativistic particles, such as Higgs boson from \sm, and members from beyond \sm, such as \dm. These produced particles cause the development of the local-thermal (for details, see~\cite{Chung:1998rq}, also of~\cite{Mukhanov:2005sc})
relativistic fluid of
the universe and consequently, raise the temperature of the universe.
Initially, due to the small value of the couplings to the inflaton~\cite{Chung:1998rq}\footnote{%
Such smallness of couplings mean the preheating particle production is not so significant~\cite{Linde:1981mu}.%
}, energy loss due to particle production is less than the loss due to expansion of the universe
~\cite{Lozanov:2019jxc}. However, the value of the Hubble parameter continues to decrease
during this reheating epoch. 
If $\G$ is the total decay width of inflaton to all other particles, then energy loss due to particle production becomes significant 
whenever ${\cal H}< \G$. If $\Trh$ be the temperature at the juncture between these two phases, then the reheating temperature is formulated as~\cite{poly}
\shila{\label{eq:definition of reheating temperature}
\Trh = \sqrt{\frac{2}{\pi}} \left(\frac{10}{g_{\star}}\right)^{1/4} \sqrt{\mp} \sqrt{\G}\,.
}

We have assumed $\gs = 106.75$
\footnote{
{
Actually, we have assumed that 
\ba
{\cal H}^2 =\frac{1}{3 \mp^2} \lt(\frac{\pi^2}{30} \lt[106.75 \,  T^4 + \frac78 (4 \, T_{\dm}^3)\rt] \rt)  \approx\frac{1}{3 \mp^2} \lt(\frac{\pi^2}{30} \lt[106.75 \, T^4 \rt] \rt) \,.\nonumber
\ea 
Here, $106.75$ is from \sm~particles
, $4$
is the degrees of freedom for fermionic \dm, and the temperature of \dm, $T_{\dm}$ is not necessarily same as $T$, the temperature of the relativistic SM fluid of the universe.
}
}
, 
which depends on the components of the relativistic thermal fluid of the universe.
Below $\Trh$, the universe behaves as if it is dominated by relativistic particles~\cite{Kolb:2003ke}.
To estimate $\Trh$ \footnote{%
There are different conventions in defining $\Trh$~as mentioned in~\cite{Garcia:2020wiy}. $\Trh$ is defined when ${\cal H}(\Trh)=\G$ or ${\cal H}(\Trh)=\frac{2}{3}\G$. These different conventions lead to different expressions of $\Trh$ in~\cite{poly} and in~\cite{Lozanov:2019jxc,Giudice:2000ex}. In this work, we follow ${\cal H}(\Trh)=\frac{2}{3}\G$.
}, we have considered that the process of particle production from inflaton is instantaneous~\cite{Chung:1998rq}.

In general, reheating is not an instantaneous process. The maximum temperature of the universe during the whole process of reheating may be many orders greater than $\Trh$.
 As reheating begins, the energy density of inflaton starts to flow into the energy density of radiation, hence increasing its temperature. However, after some initial growth of temperature, the Hubble expansion comes into play, resulting in its decline.  
The maximum temperature $\Tmax$ can be estimated as%
~\cite{Bernal:2019mhf,Giudice:2000ex,Chung:1998rq}
\shila{ \label{eq:TMAX}
\Tmax = 
\G^{1/4} \lt( \frac{60}{\gs \pi^2} \rt)^{1/4} \lt( \frac{3}{8}\rt)^{2/5} {\cal H}_I^{1/4} \mp^{1/2} \,,
}
where ${\cal H}_I$ is the value of the Hubble parameter at the beginning of reheating when no particle, including the \dm, is produced. This can be taken as 
\shila{\label{eq:HI}
{\cal H}_I 
\seq 
\bc
\sqrt{\frac{\uP (\PhZ )}{3 \mp^2}}
=
3.23 \times 10^{-10} \mp \fmI%
\,,\\ 
\sqrt{\frac{\uvp (\vpZ )}{3 \mp^2}}= 1.206 \times 10^{-10} \mp        \fmII%
\, .
\ec 
}

The simplest Lagrangian density for particle production from inflaton is assumed in Eq.~\eqref{eq:reheating lagrangian for modelI} or in Eq.~\eqref{eq:reheating lagrangian for modelII}.
In this approach, the decay width of the inflation into \sm ~Higgs field in the \mdl I, $\G_{\Ph \to \hh} $;
and in the \mdl II, $\G_{\vp \to \hh} $,
can be defined as 
\begin{align}
	\G_{\Ph (\vp) \to \hh}  &\simeq \frac{\lambda_{12}^2}{8\pi\, m_{\Phi(\vp)}}\,. %
\end{align}	
Here $h$ represents the \sm~Higgs particle. 
Likewise, we can specify the decay width of the inflaton into \dm~ particles, $\c$ as
\begin{align}
	\G_{\Ph(\vp) \to \ccp}  &\simeq \frac{y_\chi^2\, m_{\Ph (\vp)}}{8\pi}\,.
\end{align}

To satisfy present day relic density of photons and baryons, production rate of \sm~Higgs should be more than the production rate of \dm~particles during reheating, and thus, we are considering 
$\G_{\Ph(\vp)  \to \hh}  > \G_{\Ph(\vp)  \to \ccp}  $ 
such that total decay width of inflaton $\G =\G_{\Ph(\vp)  \to \ccp} + \G_{\Ph(\vp)  \to \hh} \seq \G_{\Ph(\vp)  \to \hh}$.
Here $m_\Ph$ and $m_{\vp}$ are the masses of the inflatons around $\Ph_{min}$ (in \mdl I) and around $\vp_{min}$ (in \mdl II) respectively - 
\ba
\frac{m_{\Ph (\vp)}}{\mp} = 
\bc
\lt(\mp^{-2}\dduP (\Ph) |_{\Ph=\Ph_{\rm min}} \rt)^{1/2} = 6.465 \times 10^{-9} \, \fmI  \,,\\
\lt(\mp^{-2}\dduvp (\vp) |_{\vp=\vp_{\rm min}} \rt)^{1/2} =  1.705\times 10^{-9}\,  \fmII\,. 
\ec
\ea 

Therefore, the total decay width of the inflaton 
\ba
\G = 
\bc
6.15 \times 10^6 \, \frac{\lO^2}{ \mp} \fmI \,,\\
2.33 \times 10^7 \, \frac{\lO^2}{ \mp}                    \fmII \,. %
\ec 
\ea 

\noindent Thus, from Eq.~\eqref{eq:definition of reheating temperature}, we get the reheating temperature as 
\ba\label{eq:Reheating temperature values}
\Trh =
\bc
1095.07 \, \lO \fmI \,,\\
2132.09 \lO                 \fmII\, .
\ec 
\ea 
The Eq.~\eqref{eq:Reheating temperature values} with $T \gsim 4 \text{MeV}$ puts down the lower limit on $\lO$ 
\ba\label{eq:lower limit of coupling}
\frac{\lO}{\mp}\gsim 
\bc
1.52 \times 10^{-24} \fmI \,,\\
7.82 \times 10^{-25} \fmII\, .
\ec 
\ea 
The branching ratio for the production of DM particles is given by
\shila{
\Br &= \frac{\G_{\Ph (\vp) \to \ccp} }{\G_{\Ph (\vp) \to \ccp} + \G_{\Ph (\vp) \to \hh} } \simeq  \frac{\G_{\Ph (\vp) \to \ccp} }{ \G_{\Ph (\vp) \to \hh} }   = m_{\Ph (\vp)}^2  \lt( \frac{\yc}{\lO}\rt)^2\\
&= 
\bc
4.18 \times 10^{-17} \lt(\frac{\yc}{\lO} \rt)^2 \mp^2  \fmI \,,\\
  2.91 \times 10^{-18}    \lt(\frac{\yc}{\lO} \rt)^2  \mp^2                 \fmII \, .
\ec 
} 

%
From Eq.~\eqref{eq:TMAX}, we can write
\ba\label{Eq:TmaxTrh-ratio}
\frac{\Tmax}{\Trh} = \lt(\frac{3}{8}\rt)^{2/5}  \lt( \frac{{\cal H}_I}{{\cal H}(\Trh)} \rt)^{1/4}\,,
\ea
where
\ba
{\cal H}(\Trh) = \frac{\pi}{3 \mp} \sqrt{\frac{\gs}{10}}  \Trh^2  \,.
\ea 

\begin{figure}[htp]
\centering
\begin{subfigure}{0.45\textwidth}
  \centering
 \includegraphics[width=\linewidth]{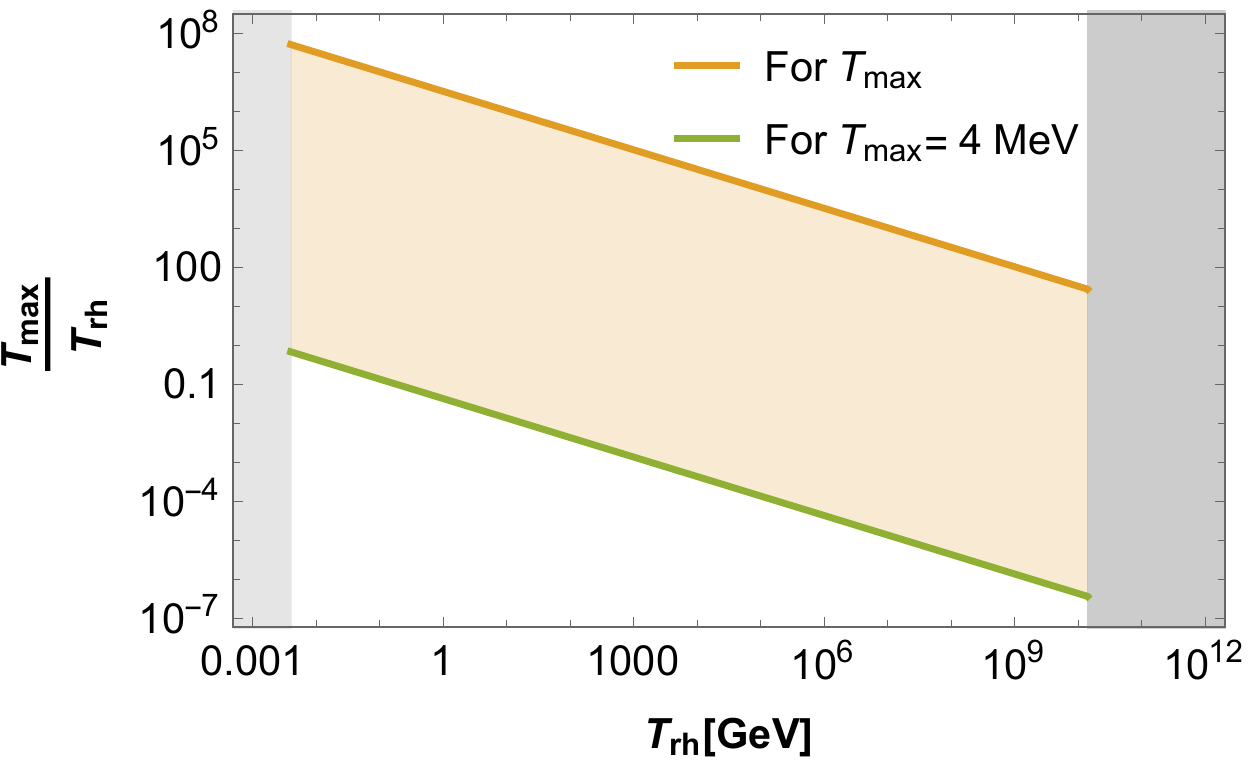}
\end{subfigure}%
\hspace{30pt}
\begin{subfigure}{.45\textwidth}
  \centering
  \includegraphics[width=\linewidth]{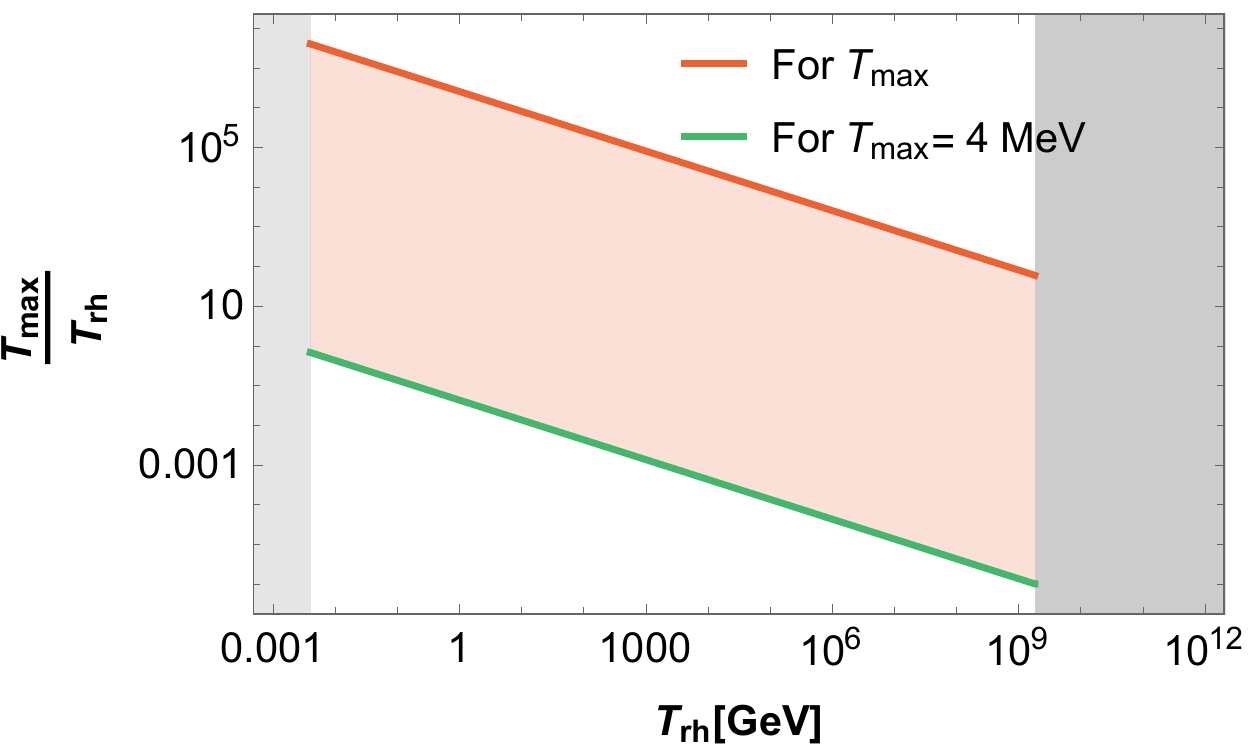}
\end{subfigure}
\vspace{-10pt}
\caption{\it  \raggedright
Allowed range (colored region) for $\Tmax/\Trh$: left panel is for the linear term inflation, where right panel is for the sextic term inflation. The green color line is for $\Tmax/\Trh$ when $\Tmax=4 \text{MeV}$. 
The gray colored region indicates the lower ($\Trh \nless 4 \text{MeV}$) and upper bound on $\Trh$. The upper bound is coming from the stability analysis (see Eq.~\eqref{eq:upper limit of l12-coupling model I} and Eq.~\eqref{eq:upper limit of l12-coupling model II})
}
\label{fig:allowed range for Tmax/Trh}
\end{figure}

The allowed ranges for $\Tmax/\Trh$ for two inflationary models are shown in Fig.~\ref{fig:allowed range for Tmax/Trh}. The upper limit for the allowed region comes from Eq.~\eqref{Eq:TmaxTrh-ratio} and the lower limit from the fact that $\Trh\gsim4\text{MeV}$ which is needed for successful Big Bang nucleosynthesis (BBN)
~\cite{Giudice:2000ex}.

\subsection{Dark Matter Production and Relic Density}
In this subsection, we estimate, following Ref.~\cite{poly}, the amount of DM produced during reheating. The Boltzmann equation for 
evolution of 
\dm~ number density, $n_\c$, is -
	\begin{equation}
	    \frac{\td n_\c}{\td t} + 3{\cal H}\,n_\c  = \gamma\,,
	\end{equation}
where $\g$ is the rate of production of \dm~ per unit volume and has the dimension of $\mp^4$.
Then the evolution equation of comoving number density, $N_\c= n_\c \ss^3$ ($\ss(t)$ is the cosmological scale factor, as mentioned earlier), 
of \dm~ particles
\ba\label{eq:Boltzaman equation for comoving number density}%
\frac{\td N_\c}{\td t} = \ss^3 \gamma  \,.
\ea 
%
During $\Trh < T< \Tmax $, the energy density of the universe is dominated by inflaton and the first Friedman equation leads to 
~\cite{poly}
\ba\label{eq:Hubble parameter during reheating}
{\cal H} = \frac{\pi}{3} \sqrt{\frac{\gs}{10}} \frac{T^4}{M_P\, \Trh^2}\,.
\ea 

Therefore, energy density of inflaton
\ba
\r_{\Ph(\vp)} = \frac{ \pi^2  \gs }{30 } \frac{T^8}{ \Trh^4} \,.
\ea

Since, during reheating, $\r_\Ph$ behaves as a non-relativistic fluids, $\r_{\Ph(\vp)} \propto \ss^{-3}$, 
the scale factor behaves as
\ba\label{eq:rheating sclae factor}
\ss\propto T^{-8/3}\,.
\ea 

Using Eq.~\eqref{eq:Hubble parameter during reheating} and \eqref{eq:rheating sclae factor} in Eq.~\eqref{eq:Boltzaman equation for comoving number density} we obtain
\ba\label{eq:dNchidT}
\frac{\td N_\c}{\td T} = - \frac{8 \mp }{\pi} \lt(\frac{10}{  \gs}\rt)^{1/2} \frac{\Trh^{10}}{T^{13}} 
\, \ss^3 (\Trh) \, \g  \,.
\ea 

\dm~$\Yc$ is defined as the ratio of the number density of \dm~to the entropy density of photons, i.e., 
$
\Yc = \frac{n_\c(T)}{s(T)}\, ,
$
where entropy density $s(T)=\frac{2 \pi^2}{45} \gss T^3 $ and $\gss$ is the effective number of  degrees of freedom of the constituents of the relativistic fluid. 
If we assume that there is no entropy generation in any cosmological process, after reheating epoch, 
then the evolution of \dm~ yield can be expressed as
	\begin{equation} \label{eq:evolution of yield}
	    \frac{\td\Yc}{\td T} = - \frac{135}{2\pi^3\, \gss} \sqrt{\frac{10}{\gs}}\, \frac{\mp}{T^6}\, \g\,.
	\end{equation}

We are assuming that
the \dm~particles, produced during reheating, were never in thermal equilibrium with the relativistic fluid of the universe. 
Those DM particles contribute to the cold dark matter (\cdm) density of the present universe.
Thus, following Table~\ref{Table:data about CDM}, present-day \cdm~ Yield~\cite{poly} is given in Eq.~\eqref{eq:present day CDM yield} 

\ba\label{eq:present day CDM yield}
 Y_{{\rm CDM},0}  = \frac{4.3. \times 10^{-10}}{\mc} \,,
\ea 
where $\mc$ is expressed in $\text{GeV}$. Now, the amount of \dm~produced during reheating through decay or via scattering in both \mdl I and \mdl II, has been estimated and compared with $ Y_{{\rm CDM},0}$ in the following part of this subsection.

\begin{table}[ht]
\begin{center}
\caption{ \it Data about CDM ($h_{\cmb}\approx0.674$)} \label{Table:data about CDM}
\begin{tabular}{ |c| c| c| }
\hline
\multicolumn{1}{|c|}{$\Omega_{\rm CDM}$} & \multicolumn{1}{c|}{ $ 0.120 \, h_{\cmb}^{-2} $} & \multirow{3}{*}{\cite{ParticleDataGroup:2020ssz}} \\
\cline{1-2}
\multicolumn{1}{|c|}{$\rho_c$} & \multicolumn{1}{c|}{$1.878 \times 10^{-29} \, h_{\cmb}^2 \, \text{g} \text{cm}^{-3}$} & \\
\cline{1-2}
\multicolumn{1}{|c|}{$s_0$} & \multicolumn{1}{c|}{$2891.2\,  (T/2.7255 \text{K})^3 \, \text{cm}^{-3}$} & \\
\hline 
\end{tabular}
\end{center}
\end{table}%
%
%
%

\subsubsection{Inflaton decay}
When \dm~ particles are produced directly from the decay of inflaton, then 
%
\ba
\g = 2 \Br \, \G \,  \frac{\r_{\Ph(\vp)}}{m_{\Ph(\vp)}} \,.
\ea 
Substituting this in Eq.~\eqref{eq:evolution of yield}, the \dm~Yield from the decay of inflaton,
%
\shila{
\Yco%
&\simeq \frac{3}{\p} \frac{\gs}{\gss} \sqrt{\frac{10}{\gs}} \frac{\mp \, \Gamma}{m_{\Ph(\vp)}\, \Trh} \text{Br}%
= \frac{3}{\p} \frac{\gs}{\gss} \sqrt{\frac{10}{\gs}} \frac{\mp }{ \Trh}  \frac{(\yc)^2}{8 \pi} \\
 &= 1.163\times 10^{-2} \mp \frac{\yc^2 }{\Trh}\,. \label{ychi0-decay-total}
}

%
Here, we assume $\gss=\gs$. Equating Eq.~\eqref{ychi0-decay-total} with Eq.~\eqref{eq:present day CDM yield}, we get 
the condition to generate the complete \cdm~energy density -%
\ba\label{eq:eq to plot 2}
\Trh  \seq  
6.49 \times 10^{25}\yc^2 \mc \,.
\ea 

\begin{figure}[htp]
\centering
\begin{subfigure}{0.45\textwidth}
  \centering
 \includegraphics[width=\linewidth]{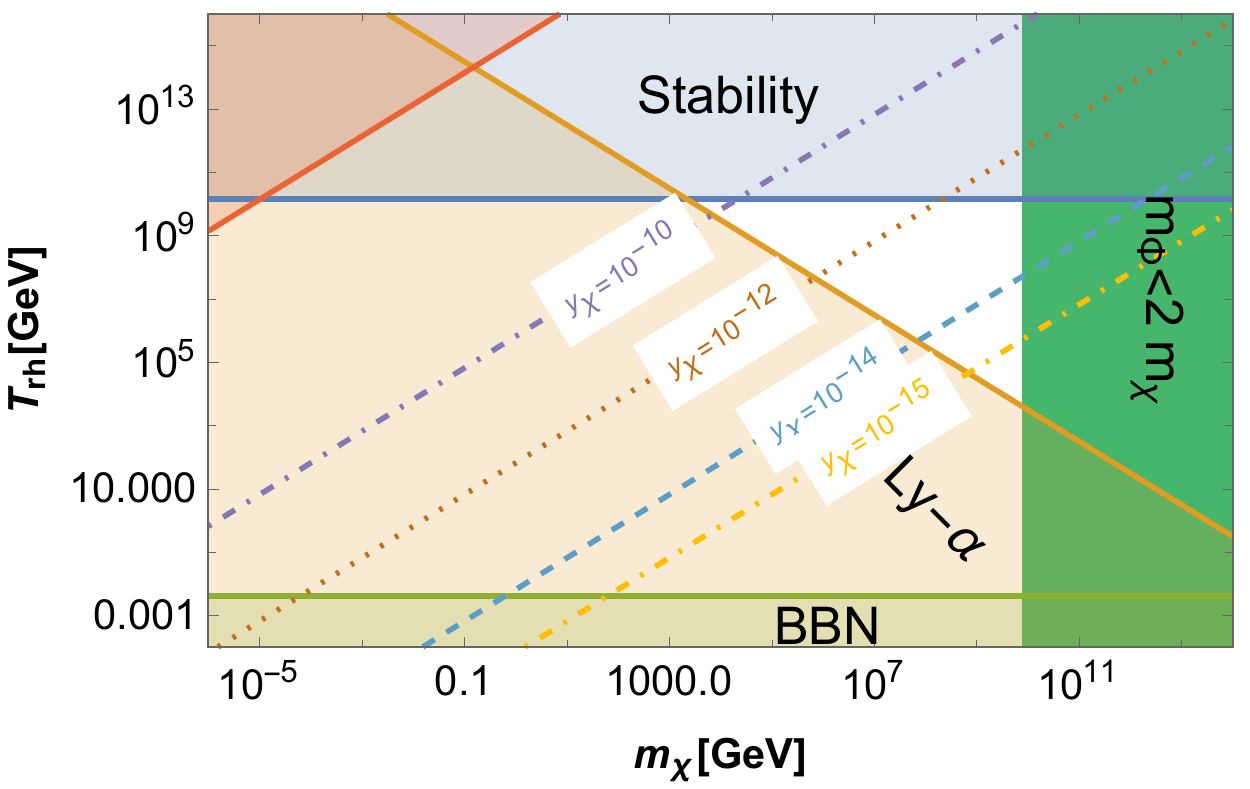}
\end{subfigure}%
\hspace{30pt}
\begin{subfigure}{.45\textwidth}
  \centering
  \includegraphics[width=\linewidth]{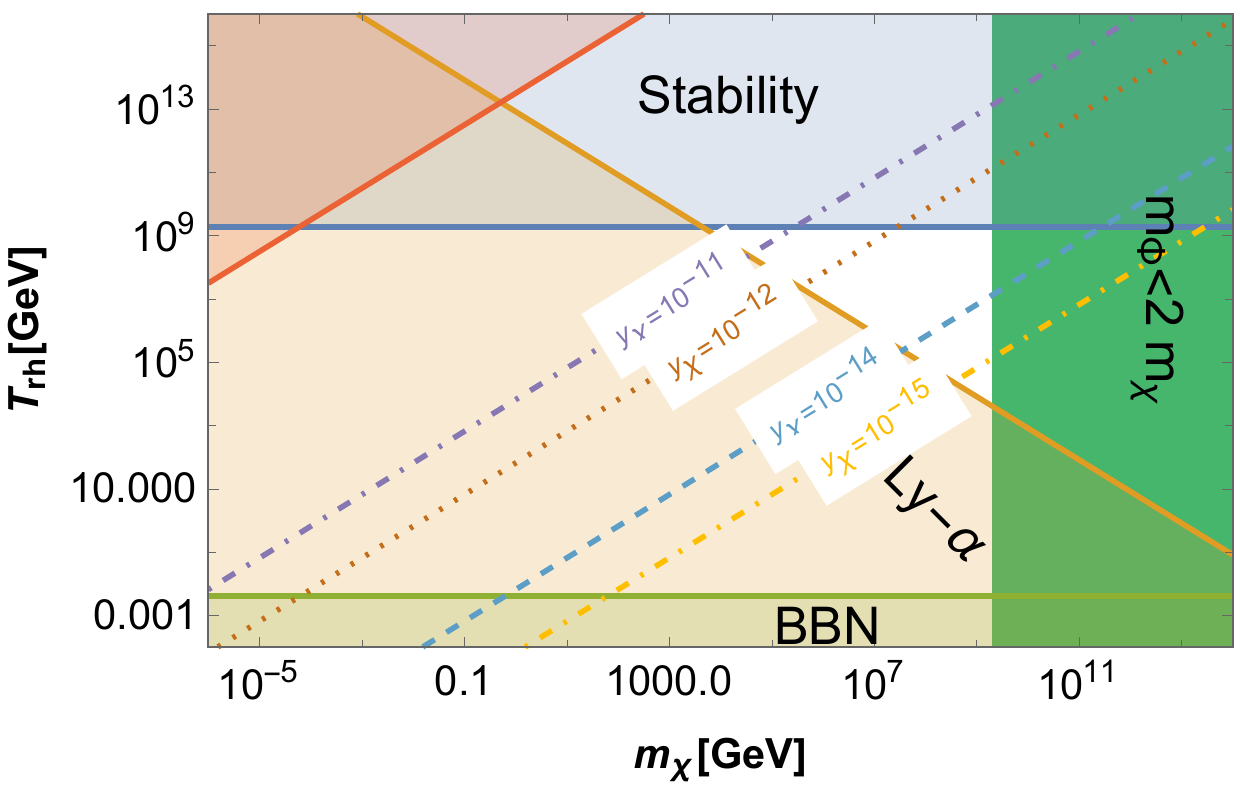}
\end{subfigure}
\vspace{-10pt}
\caption[Caption for LOF]{\it  \raggedright
The allowed region (unshaded) for the Yukawa-like coupling $\yc$ to produce the entire \cdm~of the present universe: left panel is for \mdl I inflation and right for \mdl II inflation. The constraints (colored regions) are from (a) BBN (light green colored region): $\Trh> 4 \text{MeV}$, 
(b) stability (blue colored region): $\Trh \simeq 1.388 \times 10^{10} \text{GeV}$~\fmIt or $\Trh \simeq 1.83 \times 10^{9} \text{GeV}$~\fmIIt from the upper bound of $\lO$ from Eq.~\eqref{eq:upper limit of l12-coupling model I} or Eq.~\eqref{eq:upper limit of l12-coupling model II}, (c) stability (red-colored region): from the upper bound of $\yc$ from Eq.~\eqref{eq:upper limit of yc model I} or Eq.~\eqref{eq:upper limit of yc model II}, (d) (deep green region): $\mc$ must be $<\mph/2$ (\mdl I) or $<\mvph/2$ (\mdl II), (e) (light peach-colored region):~Ly-$\a$
:~$\Trh \gsim (2  \mph)/\mc $ or $\Trh \gsim (2  \mvph)/\mc$~\cite{poly}. 
}
\label{fig:allowed range of ychi}
\end{figure}

The allowed range of the coupling $\yc$ from Eq.~\eqref{eq:eq to plot 2}, to generate the complete \cdm~density of the contemporary universe only via the decay of inflaton, is shown in Fig.~\ref{fig:allowed range of ychi}.  
From this figure, we can contemplate that the allowed range for $\yc$ and $\mc$ to construct the \cdm~density of the universe is 
$10^{-10} \gsim \yc \gsim 10^{-15}$ (for $2.5\times 10^3 \, \text{GeV} \lsim \, \mc \, \lsim 8.1\times 10^{9} \, \text{GeV}$ in \mdl I) and  $10^{-11} \gsim \yc \gsim 10^{-15}$ (for $8.4\times 10^3 \, \text{GeV} \lsim \, \mc \, \lsim 2\times 10^{9} \, \text{GeV}$ in \mdl II).

In Fig.~\ref{fig:allowed range of ychi} we show the bounds from Lyman-$\a$ forest which is a series of absorption lines on the received spectra of distant (at high redshift) quasars or bright galaxies.  The accountability of these absorption lines, corresponding to Lyman-$\a$-electron-transition, mainly goes to neutral hydrogen present in the intergalactic medium and also to the change of the wavelength due to redshift. 
Since, hydrogen in the intergalactic medium \textit{traces matter power spectrum}, it is possible to estimate the lower limit of the mass of warm dark matter from the data about small scale structure obtained from Lyman-$\a$ forest flux power spectrum~\cite{Berbig:2022nre}. The Lyman-$\a$ bound used in this work is drawn with the assumption that momentum of warm dark matter (WDM) has been only redshifted since its origin. Even though the bound has been shown to have modified to some extent for \dm-\sm~interaction~\cite{Ghosh:2022hen}, it does not affect the present analysis, as \dm-\sm~interaction strength is negligible here. Moreover, to derive the Lyman-$\a$ bound it is also considered that the lower bound on the mass of warm dark matter particle $m_{WDM}\gsim 3.5 \text{keV}$~\cite{poly}. 
For thermal warm dark matter, $m_{WDM}\gsim 5.3 \text{keV}$ at $95\%$ CL~\cite{Viel:2013fqw,Palanque-Delabrouille:2019iyz} and it may change to $m_{WDM}\gsim 1.9 \text{keV}$~\cite{Garzilli:2019qki} if the uncertainties in the thermal history of the universe is brought under consideration. Further studies claim that the bound on other kinds of FIMPs is $m_{WDM}\gsim (4-16) \text{keV}$~\cite{Bae:2017dpt,Murgia:2017lwo,Heeck:2017xbu,Boulebnane:2017fxw,Baldes:2020nuv,Ballesteros:2020adh,DEramo:2020gpr,Decant:2021mhj}.


\subsubsection{Inflaton Scattering}
\label{Inflaton-Scattering}
\begin{figure}[htp]%
\centering
\begin{subfigure}{0.45\textwidth}
  \centering
 \includegraphics[width=\linewidth]{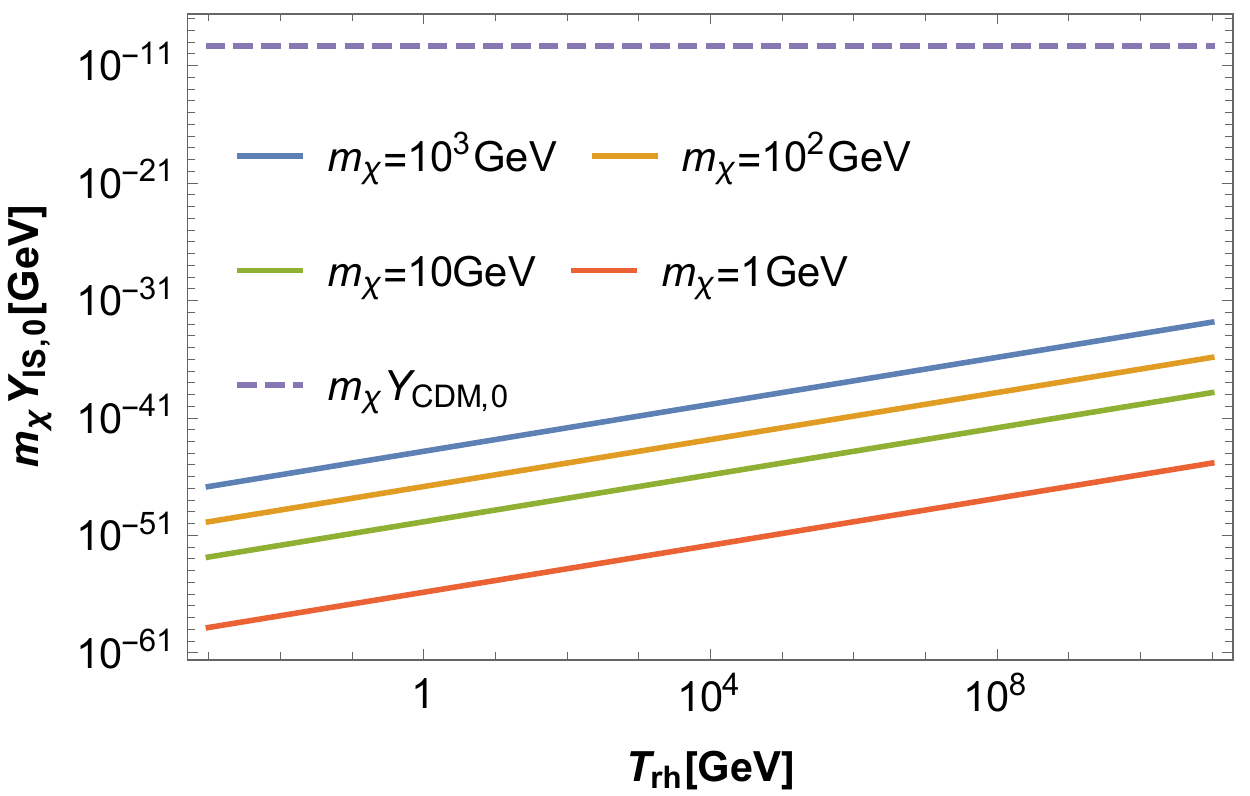}
\end{subfigure}%
\hspace{30pt}
\begin{subfigure}{.45\textwidth}
  \centering
  \includegraphics[width=\linewidth]{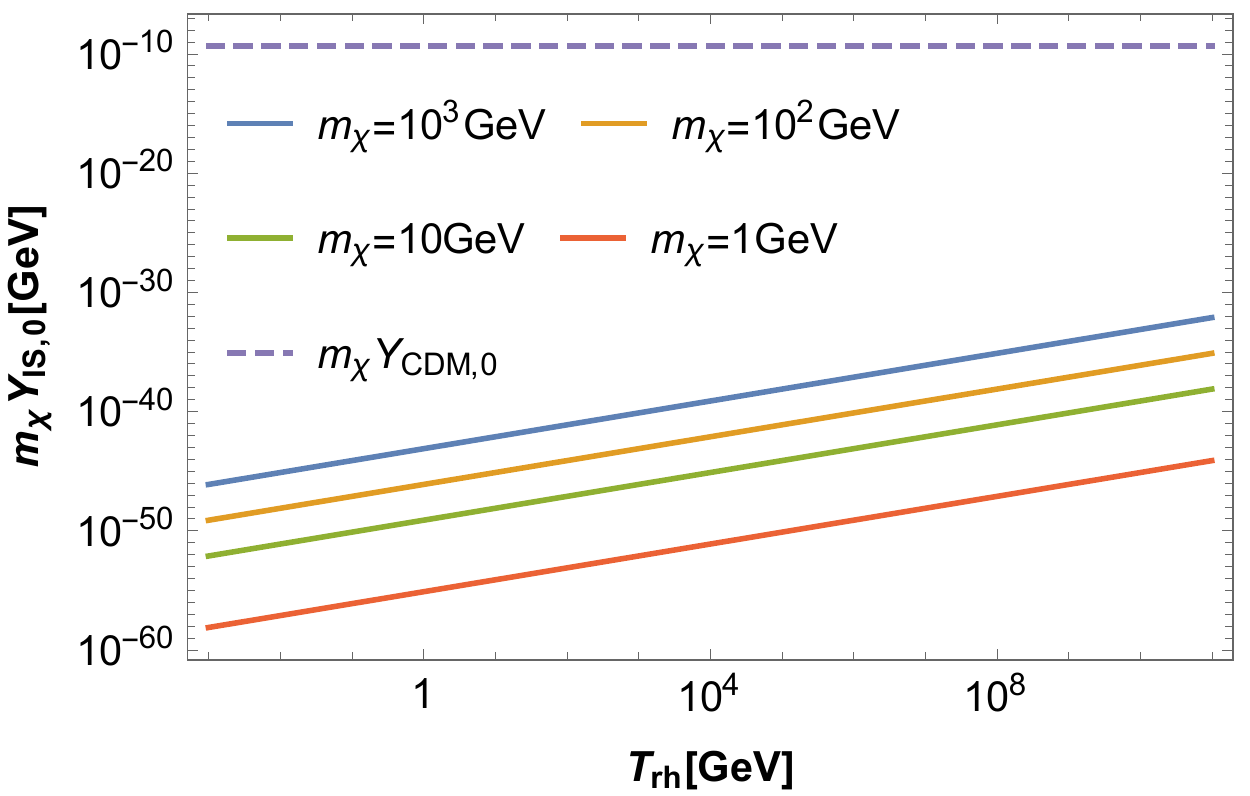}
\end{subfigure}
\vspace{-10pt}
\caption{\it  \raggedright
$\mc \times$Yield of \dm~ produced from the 2-to-2 scattering with graviton as mediator for different values of $\mc$. Left panel is for \mdl I and right panel for \mdl II inflation.
}
\label{fig:mc YIS0 - from inflaton scattering via graviton}
\end{figure}

In this work, we consider the important 2-to-2 scattering processes for the \dm~ production, as mentioned in~\cite{poly}.
When graviton acts as the mediator for the production of \dm~ particles from non-relativistic inflaton via 2-to-2 scattering, then~\cite{poly}
    \begin{equation}
        \gamma = \frac{\pi^3\, \gs^2}{3686400} \frac{T^{16}}{M_P^4\, \Trh^8} \frac{m_{\chi}^2}{m_{\Ph(\vp)}^2} \left(1-\frac{m_{\chi}^2}{m_{\Ph(\vp)}^2}\right)^{3/2}\,,
    \end{equation}
and thus Eq.~\eqref{eq:evolution of yield} prompts to the \dm~yield 
	\begin{align} \label{eq:yield-DM-scattering-inflaon-graviton}
	    Y_{IS,0} &\simeq \frac{\gs^2}{81920\gss} \sqrt{\frac{10}{\gs}} \left(\frac{\Trh}{M_P}\right)^3 \left[\left(\frac{\Tmax}{\Trh}\right)^4 - 1 \right] \frac{m_{\chi}^2}{m_{\Ph(\vp)}^2} \left(1-\frac{m_{\chi}^2}{m_{\Ph(\vp)}^2}\right)^{3/2}\,.
    \end{align}
In Fig.~\ref{fig:mc YIS0 - from inflaton scattering via graviton}, $Y_{IS,0}$ {(actually $\mc Y_{IS,0}$ with $\mc Y_{{\rm CDM},0}$)}  is compared with $Y_{{\rm CDM},0}$ for different $\mc$ as a function of $\Trh$. Hence, it is shown there that the yield  of \dm~ produced via scattering (Eq.~\eqref{eq:yield-DM-scattering-inflaon-graviton}) is not significant compared to the present \cdm~density.

\subsubsection{\sm~ Scattering}
\label{se:SM-Scattering}
\dm~particles can also be produced from the scattering of \sm~particles via graviton mediation. In that case,
\begin{equation}
	\gamma = \alpha\, \frac{T^8}{M_P^4}\,,
\end{equation}
where $\alpha\seq 1.1 \times 10^{-3}$. 
Due to the presence of $\mp^4$ in the denominator, it is expected that the production of \dm~through this process is less compared to previous ones. 
Substituting this in Eq.~\eqref{eq:evolution of yield}, and integrating, we get the \dm~Yield, $Y_{\sm g,0}$ produced through this scattering channel as-
	\begin{equation} \label{smg1}
	Y_{\sm g,0} = 
	\begin{cases}
	\frac{45 \alpha}{2\pi^3 \gss} \sqrt{\frac{10}{\gs}} \left(\frac{\Trh}{M_P}\right)^3, \qquad \text{ for } m_\chi \ll \Trh\,,\\
	\frac{45 \alpha}{2\pi^3 \gss} \sqrt{\frac{10}{\gs}} \frac{\Trh^7}{M_P^3\, m_\chi^4}\,, \qquad \text{ for } \Tmax \gg m_\chi \gg \Trh\,.
	\end{cases}
	\end{equation}

For this process, $Y_{\sm g,0}\sim 10^{-52}$ when $\mc \ll \Trh=10^3\text{GeV}$ and $Y_{\sm g,0}\sim 10^{-76}$ when $\mc= 100 \text{GeV}, \,  \Trh=0.1\text{GeV}$ for both inflationary models we considered.

When inflaton acts as mediator for the production of \dm~from  2-to-2 scattering of \sm~particles,
\ba
		\gamma(T) \seq \frac{\yc^2\, \lambda_{12}^2 }{2 \pi^5} \frac{T^6}{m_{\Ph(\vp)}^4}\,,
\ea 
and production of \dm~(Yield) only through that channel results in 

	\begin{equation} 
	Y_{\sm i,0} \simeq \frac{135\,  y^2_{\chi}\, \lambda_{12}^2}{4 \pi^8\, \gss} \sqrt{\frac{10}{g_\star}}\, \frac{M_P\, \Trh}{m_{\Ph(\vp)}^4}\, , \qquad \text{ for } \Trh \ll m_{\Ph(\vp)}, \Trh > T \,.
	\end{equation}

$Y_{\sm i,0}\sim 10^{-60}$ ($\sim 10^{-62}$) for $\Trh\sim 10^{5} \text{GeV} \simeq 10^{-5} \mph$ ($\mvph$) for $\gs=\gss=106.75$, $\lO\sim 10^{-12}$ ($10^{-13}$) and $\yc \sim 10^{-6}$ ($10^{-7}$). 
Therefore, the \dm~produced from 2-to-2 scattering during reheating is 
insignificant compared 
in comparison to total \cdm~density of the universe.

\section{Conclusions and Discussion}
\label{Sec:Conclusions and Discussion}
We investigated a simple possibility of a scalar inflaton and a fermionic particle behaving as the CDM, the latter being produced from the inflaton decay. Satisfying the correct relic density of DM, we observed the salient features of our analysis:
\begin{itemize}
\item We explored two models of polynomial potential, each of which has an inflection point, leading to the slow roll single field cosmological inflation. The potential of \mdl I contains a term proportional to the linear power of inflaton (see Eq.~\eqref{eq:inflation potential of model I}). On the other hand, the potential of \mdl II (Eq.~\eqref{eq:inflation potential of model II}) is symmetric under the transformation of $\vp\to -\vp$. 
 
\item  Assuming the near-inflection point inflationary scenarios for both models, we estimated the coefficients of the potentials of both models which satisfies the CMB constraints. We found $n_s\sim 0.96$, $r\sim 10^{-12}, \a_s\sim 10^{-3}$, and $\b_s\sim 10^{-8}$  (see Table~\ref{Tab:Model I benchmark values} and Table~\ref{Tab:Model II benchmark values}). 

\item  We considered the production of non-thermal vector-like fermionic DM particles, $\chi$, from the decaying inflaton during reheating. The rate of production of DM through this decay depends on the reheating temperature; as the temperature of the relativistic fluid of the universe during reheating increases, production rate of \dm~decreases  (Eq.~\eqref{eq:evolution of yield}). The allowed range for the ratio of the maximum temperature $\Tmax$ to the reheating temperature $\Trh$ during that epoch, $\Tmax/\Trh$, is shown in Fig.~\ref{fig:allowed range for Tmax/Trh}.  The ratio can be as high as ${\cal O}(10^{7})$ for $\Trh=4\text{MeV}$. The allowed region of $\Tmax/\Trh$ depends on the inflection point (see Eq.~\eqref{eq:TMAX} and Eq.~\eqref{eq:HI}).
Since we choose the \cmb~scale near the inflection point, the choice of inflection point controls the \cmb~observables, e.g. $n_s$ and $r$ on one hand, and controls the production regimes (via T$_{max}$) of DM and consequently DM relic on the other hand.

\item We also assumed that inflaton decays to \sm~Higgs ($H$) together with $\chi$. 
From stability analysis of the inflation-potential in Fig.~\ref{fig:stability_analysis_plot_for_model_I} and Fig.~\ref{fig:stability_analysis_plot_for_model_II}, we demarcated that the upper bound of the couplings for two decay channels are $\lO/\mp \lsim {\cal O}(10^{-12})$ and $\yc \lsim {\cal O}(10^{-6})$. The former upper bound outlines the  maximum permissible value of $\Trh$ for our models. 

\item  From Fig.~\ref{fig:allowed range of ychi}, we can conclude that $\chi$ produced only through the decay of inflaton may explain the total  density of \cdm~of the present universe if $10^{-10} \gsim \yc \gsim 10^{-15}$ (for $2.5\times 10^3 \, \text{GeV} \lsim \, \mc \, \lsim 8.1\times 10^{9} \, \text{GeV}$ in \mdl I) and  $10^{-11} \gsim \yc \gsim 10^{-15}$ (for $8.4\times 10^3 \, \text{GeV} \lsim \, \mc \, \lsim 2\times 10^{9} \, \text{GeV}$ in \mdl II).
Actually, Fig.~\ref{fig:allowed range of ychi} depicts the allowed region in $\Trh-\mc$ space for two models of potential we have considered and
the
constraints on that space are coming from bound on $\Trh$ from \bbn, radiative stability analysis of the potential for slow roll inflation, Ly-$\a$ bound, and the maximum possible value of $\mc$ for the effective mass of the inflaton.
\\

\item  $\chi$ can also be produced from 2-to-2 scattering of either \sm~ particles or inflatons. Among them, the promising ones are – from the scattering of inflatons with graviton as the mediator and from the scattering of two \sm~  particles with either graviton or inflaton as the mediator. In Fig.~\ref{fig:mc YIS0 - from inflaton scattering via graviton} we showed that $\Yc$ produced through 2-to-2 scattering of inflaton with graviton as mediator, is {more} than the \dm~ production via other scattering channels, and it is $ Y_{IS,0}\sim  {\cal O}( 10^{-36}$) for $\Trh=10^8~\text{GeV}, \mc= 10^3~\text{GeV}$.
But, $ Y_{IS,0}$ produced through this channel is much less than $Y_{{\rm CDM},0}$ and thus $\chi$ produced through 2-to-2 scattering channels can  contribute only a negligible fraction of $Y_{{\rm CDM},0}$. 
\end{itemize}

In a nutshell, in this article, just by extending the standard model with two degrees of freedom: a real scalar inflaton, and a fermionic DM,  we have been able to demonstrate that one may explain the tiny temperature fluctuations as seen in the CMBR as well as address the dark matter puzzle of the Universe, in terms of its possible origin as well as satisfying the combined constraints coming from different observations. 

Future measurements of the CMB from experiments like CMB-S4, SPTpol, LiteBIRD, and CMB-Bharat~\cite{CMB-S4:2020lpa,Hazumi:2019lys,Adak:2021lbu,SPT:2019nip} and other such experiments \cite{POLARBEAR:2015ixw,ACT:2020gnv,Harrington:2016jrz,LSPE:2020uos, Mennella:2019cwk, SimonsObservatory:2018koc,SPIDER:2021ncy}
 will further be able to test the simple models we have presented if BB-modes are detected and the scale of inflation is measured. If not, this can then give us a hint towards exploring possible avenues of verifying the models, from both theoretical and observational points of view. Analysis of the model in terms of scalar perturbations will be an interesting direction to build upon in future as such inflection-points are known to give significant production of PBHs possibly with extensions to the simple models presented in this paper and scalar-induced Gravitational Waves~\cite{Chatterjee:2017hru,Choudhury:2013woa,Bhaumik:2022pil,Domenech:2021ztg} all of which are targets of future GW observatories and PBH observations. Such computations along with non-Gaussianities will be very interesting probes of these models and complementary signatures of non-thermal production of DM and will be taken up in future publications.  

\medskip

\section*{Acknowledgement}
Authors acknowledge communications with Alexander Dmitrievich Dolgov, Anupam Mazumdar, Arindam Chatterjee, Yong Xu and Raghavan Rangarajan.
~Work of Shiladitya Porey is funded by RSF Grant 19-42-02004. Supratik Pal thanks Department of Science and Technology, Govt. of India
for partial support through Grant No. NMICPS/006/MD/2020-21.

\medskip

\end{document}